\documentclass[12pt,draftclsnofoot,onecolumn]{IEEEtran}

\IEEEoverridecommandlockouts
\usepackage{cite}
\usepackage{indentfirst,flushend}
\usepackage[justification=centering]{caption}
\usepackage{amssymb,amscd,amsmath,amsthm,amsfonts,mathrsfs,times,bm,bbm}
\usepackage[colorlinks,linkcolor=black,anchorcolor=black,citecolor=black]{hyperref}
\usepackage{latexsym}
\usepackage{graphicx}
\usepackage{subfigure}
\usepackage[usenames,dvipsnames]{color}
\usepackage{cases}
\usepackage{algorithm}
\usepackage{algorithmic}
\usepackage{clrscode}
\usepackage{psfrag,stfloats,nicefrac}
\usepackage{supertabular}
\usepackage{makecell}
\usepackage{booktabs}
\usepackage{array}
\usepackage{multirow}
\usepackage{threeparttable}
\usepackage{color,soul,framed}
\allowdisplaybreaks[4]
\newtheorem{lemma}{Lemma}

\newcommand{\rmnum}[1]{\uppercase\expandafter{\romannumeral #1\relax}}
\newcommand{\blue}[1]{\color{RoyalBlue}{#1}}
\newcommand{\textred}[1]{\textcolor{red}{#1}} 

\begin{document}

\title{Joint User Scheduling and Beamforming Design for Multiuser MISO Downlink Systems}
\author{
Shiwen~He,~\IEEEmembership{Member,~IEEE},~Jun~Yuan,~Zhenyu An,~\IEEEmembership{Student Member,~IEEE},~Wei\\ Huang,~\IEEEmembership{Member,~IEEE},Yongming Huang,~\IEEEmembership{Senior Member,~IEEE},~and Yaoxue~Zhang,~\IEEEmembership{Senior Member,~IEEE}
\thanks{S. He and J. Yuan are with the School of Computer Science and Engineering, Central South University, Changsha 410083, China. S. He is also with the National Mobile Communications Research Laboratory, Southeast University, and the Purple Mountain Laboratories, Nanjing 210096, China. (email: \{shiwen.he.hn, yuanjun\}@csu.edu.cn). }
\thanks{Z. An is with the Purple Mountain Laboratories, Nanjing 210096, China. (email: anzhenyu@pmlabs.com.cn). }
\thanks{W. Huang is with the School of Computer Science and Information Engineering, Hefei University of Technology, Hefei 230601, China. (email: huangwei@hfut.edu.cn)}
\thanks{Y. Huang is with the National Mobile Communications Research Laboratory, School of Informatiecience and Engineering, Southeast University, Nanjing 210096, China. He is also with the Purple Mountain Laboratories, Nanjing 210096, China. (email: huangym@seu.edu.cn). }
\thanks{Y. Zhang is with the Department of Computer Science and Technology, Tsinghua University, Beijing 100084, China. (email: zhangyx@tsinghua.edu.cn)}
}
\maketitle
\vspace{-.6 in}

\begin{abstract}
In multiuser communication systems, user scheduling and beamforming (US-BF) design are two fundamental problems that are usually studied separately in the existing literature. In this work, we focus on the joint US-BF design with the goal of maximizing the set cardinality of scheduled users, which is computationally challenging due to the non-convex objective function and the coupled constraints with discrete-continuous variables. To tackle these difficulties, a successive convex approximation based US-BF (SCA-USBF) optimization algorithm is firstly proposed. Then, inspired by wireless intelligent communication, a graph neural network based joint US-BF (J-USBF) learning algorithm is developed by combining the joint US and power allocation network model with the BF analytical solution. The effectiveness of SCA-USBF and J-USBF is verified by various numerical results, the latter achieves close performance and higher computational efficiency. Furthermore, the proposed J-USBF also enjoys the generalizability in dynamic wireless network scenarios.
\end{abstract}

\section{Introduction}
With the explosive growth of Internet of Things (IoT) devices, wireless communication networks (WCNs) are increasingly facing the challenge of allocating finite transmit power and bandwidth for system utility maximization~\cite{xu2021survey}. Accordingly, one needs to design advanced radio resource management schemes to serve numerous wireless access devices. Massive multiple-input multiple-output (MIMO) and multiuser transmission are two key enablers for supporting larger-scale connection in future WCNs~\cite{he2021survey}. Therefore, some works have been carried on researching the beamforming design (BF)~\cite{he2015energy}, power allocation (PA)~\cite{yu2020power}, and user scheduling (US)~\cite{ammar2021distributed}, etc.

Generally speaking, US and BF (US-BF) design are two fundamental problems in multiuser WCNs, \textred{which are implemented at the media access control layer \cite{dimic2005on} and the physical layer \cite{zhang2009networked}, respectively. Unfortunately, these two issues are always coupled, which is difficult to be solved.} Therefore, they are generally investigated separately in the existing literature, such as BF design with a given user set~\cite{shi2011iteratively} or US optimization combined with PA (US-PA)\cite{dong2019energy}. \textred{For example, the authors of~\cite{yu2007transmitter} and~\cite{huh2012network} only consider the BF problem, where the uplink-downlink duality theory is adopted for tackling the non-convex problem of transceivers design. The authors of \cite{huang2020hybrid} and \cite{huang2021multi-hop} also solve the BF problem for RIS-empowered Terahertz communications with deep reinforcement learning methods. To further improve the performance of WCNs, cross-layer design is increasingly becoming popular~\cite{fu2014a}. The authors of~\cite{yoo2006on} investigate the US-BF problem by sequentially performing the semi-orthogonal user selection (SUS) algorithm for US optimization and the zero-forcing BF (ZFBF) algorithm for BF design. The authors of \cite{chen2017low} propose a low complexity US-BF scheme for 5G MIMO nonorthogonal multiple-access systems, but the non-convex problem is separated by tracking two subproblems, namely, BF scheme and greedy min-power US scheme, instead of jointly solving them. The authors of~\cite{zhang2017sum-rate} also discuss cross layer optimization with statistical channel information for massive MIMO scenario, by tackling US and BF individually.}

Meanwhile, the existing researches on coordinated multiuser communication are mainly based on the conventional Shannon theory~\cite{shannon1948mathematical}, which assumes that the communication capacity has extremely low decoding error probability with enough long blocklength transmission. However, in the ultra-reliable low latency communication (URLLC) senarios, such as factory automation and remote surgery, this condition with the long blocklength transmission may not be satisfied~\cite{nasir2020resource}. To take the impact of finite blocklength transmission into account, the achievable rate has been expressed as a complicated function composed of the received signal-to-noise (SNR), the blocklength, and the decoding error probability, which is smaller than the Shannon rate~\cite{polyanskiy2010channel}. Consequently, the optimization problem in scenarios with finite blocklength transmission is more challenging~\cite{he2020beamforming}. In order to solve the problem of interest, the algorithms designed in the aforementioned references are mainly based on the convex optimization theory~\cite{bertsekas2003convex}. However, such model-driven optimization algorithms usually suffer from a high computational complexity, which may restrict their practical application ability in WCNs.

Recently, deep neural networks (DNNs) have emerged as an effective tool to solve such challenging radio resource management problems in WCNs~\cite{she2021tutorial}. Different from the model-driven optimization algorithms running independently for each instance, DNNs are trained with numerous data to learn the mapping between radio resource optimization policies and WCN environments. Hence, the main computational cost of DNNs is shifted into the offline training stage, and only simple mathematical operations are needed in the online optimization stage. The work in~\cite{li2021multicell} shows that DNNs could achieve competitive performance with lower computational complexity than existing model-driven optimization algorithms. A similar conclusion has been demonstrated in~\cite{xia2020deep}, where DNNs are used for BF design of multiuser multiple-input single-output (MISO) downlink systems, but the size of the considered problem is rather small. \textred{The authors of \cite{kaushik2021} regard resource allocation problems in the field of wireless communications as the generalized assignment problems (GAP), and propose a novel deep unsupervised learning approach to solve GAP in a time-efficient manner. The authors of \cite{liang2020towards} focus on solving PA problem via ensembling several deep neural networks. This is also an unsupervised approach and achieves competitive results compared with conventional methods. However, the core network is specifically designed for power control problem and it could not be extended for US.} In addition, these DNN-based architectures~\cite{liang2020towards,kaushik2021,li2021multicell,xia2020deep} are mainly inherited from image processing tasks and not tailored to radio resource management problems, especially the fact that they fail to exploit the prior topology knowledge in WCNs. The numerical results obtained in~\cite{chen2021gnn} illustrated that the performance of DNNs degrades dramatically with increasing WCN size.

To achieve a better scalability of learning-based radio resource management, a potential approach is to incorporate the network topology into the learning of neural networks, namely graph neural networks (GNNs)~\cite{he2021overview}. For instance, the authors of~\cite{cui2019spatial} combined DNNs with the geographic location of transceivers, and thereby proposed a spatial convolution model for wireless link scheduling problems with hundreds of nodes. The authors of~\cite{eisen2020optimal} proposed a random edge graph neural network (REGNN) for PA optimization on graphs formed by the interference links within WCNs. The work in~\cite{shen2019graph} demonstrates that GNNs are insensitive to the permutation of data, such as channel state information (CSI). Further, this work was extended in~\cite{shen2020graph} to solve both PA and BF problems via message passing graph neural networks (MPGNNs), which have the ability to generalize to large-scale problems while enjoying a high computational efficiency. However, their proposed designs in~\cite{shen2019graph,shen2020graph} only investigated the continuous optimization problems with simple constraints. The discrete optimization problems with complicated constraints are still an opening issue and need to be further considered. Fortunately, the application of primal-dual learning in~\cite{he2021gblinks} provides an effective way to solve the complicated constrained radio resource management problems.

\textred{Based on the above considerations, this work studies the joint US-BF optimization problem in the multiuser MISO downlink system. Unlike the conventional methods, the US-BF design will be simultaneously achieved via solving a single optimization problem, instead of different problems. Moreover, to improve the computational efficiency and utilize network historical data information, we propose a GNN-based Joint US-BF (J-USBF) learning algorithm. The main contributions and advantages of this work are summarized as follows:}

\begin{itemize}
\item \textred{A joint US-BF optimization problem for multiuser MISO downlink systems is formulated with the goal of maximizing the number of scheduled users subject to user rate and base station (BS) power constraints. To solve this discrete-continuous variables optimization problem, a SCA-based US-BF (SCA-USBF) algorithm is firstly designed to pave the way for the J-USBF algorithm.}
\item \textred{A J-USBF learning algorithm is developed by combining the joint user scheduling and power allocation network (JEEPON) model with the BF analytical solution. In particular, we first formulate the investigated problem as a graph optimization problem through wireless graph representation, then design a GNN-based JEEPON model to learn the US-PA strategy on graphs, and utilize the BF analytical solution to achieve joint US-BF design. Meanwhile, a primal-dual learning framework is developed to train JEEPON in an unsupervised manner.}
\item Finally, numerical results is conducted to validate the effectiveness of the proposed algorithms. Compared with the SCA-USBF algorithm, the J-USBF learning algorithm achieves close performance and higher computational efficiency, and enjoys the generalizability in dynamic WCN scenarios.
\end{itemize}

The remainder of this paper is organized as follows. Section~\rmnum{2} introduces a challenging radio resource management problem in the multiuser MISO downlink system. Section~\rmnum{3} proposes the SCA-USBF for solving the investigated problem. Section~\rmnum{4} designs the JEEPON and provides a primal-dual learning framework to train it in an unsupervised manner. Numerical results are presented in Section~\rmnum{5}. Finally, conclusions are drawn in Section~\rmnum{6}.

\textbf{\textcolor{black}{$\mathbf{\mathit{Notations}}$}}: Throughout this paper, lowercase and uppercase letters (such as $a$ and $A$) represent scalars, while the bold counterparts $\mathbf{a}$ and $\mathbf{A}$ represent vectors and matrices, respectively. $\left|\cdot\right|$ indicates the absolute value of a complex scalar or the cardinality of a set. $\left\|\cdot\right\|_{0}$, $\left\|\cdot\right\|_{1}$, and $\left\|\cdot\right\|_{2}$ denote the $\ell_{0}$-norm, $\ell_{1}$-norm, and $\ell_{2}$-norm, respectively. The superscripts $(\cdot)^{T}$, $(\cdot)^{H}$, and $(\cdot)^{-1}$ denote the transpose, conjugate transpose, and inverse of a matrix, respectively. $\mathbb{R}$, $\mathbb{R}^{+}$, and $\mathbb{C}$ are the sets of real, non-negative real, and complex numbers, respectively. Finally, $\mathbb{R}^{M\times1}$ and $\mathbb{C}^{M\times1}$ represent $M$-dimensional real and complex column vectors, respectively.

\section{System Model and Problem Formulation}
In this work, we consider a multiuser MISO downlink system with taking the reliable and delivery latency into account, where a BS with $N$ antennas serves $K$ single-antenna users\footnote{\textred{Since the complexity of discussed problem, the single-cell scenario is considered in this paper. Research on more complex scenario with multi-cells will be discussed in future work, where inter-cell interference should be considered.}}. For simplicity, let $\mathcal{K}=\{1,2,\cdots,K\}$ and $\mathcal{S}=\{1,2,\cdots,K^{\ast}\}\subseteq\mathcal{K}$ be the set of candidate users and scheduled users, respectively, where $K^{\ast}\leq{K}$. The channel between user $k$ and the BS is denoted as $\mathbf{h}_{k}\in\mathbb{C}^{N\times1}$. Let $p_{k}\geq{0}$ and $\mathbf{w}_{k}\in\mathbb{C}^{N\times1}$ represent the transmit power and unit-norm BF vector used by the BS for user $k$, respectively. Thus, the received signal at user $k$ is given by
\begin{equation}\label{Eq.(01)}
y_{k}=\sum\limits_{l\in\mathcal{S}}\sqrt{p_{l}}\mathbf{h}_{k}^{H}\mathbf{w}_{l}s_{l}+n_{k},
\end{equation}
where $s_{l}$ is the normalized data symbol intended for the $l$-th user, and $n_{k}\sim\mathcal{CN}(0,\sigma_{k}^{2})$ denotes the additive Gaussian white noise at user $k$ with zero mean and variance $\sigma_{k}^{2}$. For notational convenience, we define $\overline{\mathbf{h}}_{k}=\frac{\mathbf{h}_{k}}{\sigma_{k}}$ and the downlink signal-to-interference-plus-noise ratio (SINR) of user $k$ as
\begin{equation}\label{Eq.(02)}
\overrightarrow{\gamma}_{k}=\frac{p_{k}\left|\overline{\mathbf{h}}_{k}^{H}\mathbf{w}_{k}\right|^{2}}
{\sum\limits_{l\neq k,l\in\mathcal{S}}p_{l}\left|\overline{\mathbf{h}}_{k}^{H}\mathbf{w}_{l}\right|^{2}+1}.
\end{equation}

To satisfy the extreme requirements of delay, finite blocklength transmission regime is adopted in this paper. The results in~\cite{polyanskiy2010channel} show that the achievable rate is not only a function of the received SNR (or SINR), but also the decoding error probability $\epsilon$ and the transmission finite blocklength $n$. Accordingly, the achievable rate of user $k$ with finite blocklength transmission is given by\footnote{The proposed algorithms is also suitable for solving similar optimization problems, where the user rate is based on Shannon capacity formula.}
\begin{equation}\label{Eq.(03)}
R(\overrightarrow{\gamma}_{k})=C(\overrightarrow{\gamma}_{k})-\vartheta\sqrt{V(\overrightarrow{\gamma}_{k})},
\end{equation}
where $C(\overrightarrow{\gamma}_{k})=\ln(1+\overrightarrow{\gamma}_{k})$ denotes the Shannon capacity, $\vartheta=\frac{Q^{-1}(\epsilon)}{\sqrt{n}}$, $Q^{-1}(\cdot)$ is the inverse of Gaussian Q-function $Q(x)=\frac{1}{\sqrt{2\pi}}\int_{x}^{\infty}\mathrm{exp}(-\frac{t^{2}}{2})dt$, and $V(\overrightarrow{\gamma}_{k})$ denotes the channel dispersion, which is defined as
\begin{equation}\label{Eq.(04)}
V(\overrightarrow{\gamma}_{k})=1-\frac{1}{(1+\overrightarrow{\gamma}_{k})^{2}}.
\end{equation}

The target of this work is to maximize the number of users belonging to the scheduled user set $\mathcal{S}\subseteq\mathcal{K}$ subject to the constraints of per-user minimum rate requirement and BS maximum power budget. Specifically, one needs to carefully select the scheduled user set $\mathcal{S}$, and design BF vectors with reasonable transmit power\footnote{For the ultra-dense or large-scale connective URLLC scenario, it may be a better choice to schedule as many users as possible while satisfying reliability and latency requirements. Accordingly, we aim to maximize the set cardinality of scheduled users in this work.}. To this end, the joint US-BF optimization problem is formulated as follows\footnote{\textred{In our experiment, we obtain perfect CSI via link level simulation. However, it is indeed hard to estimate CSI in the real communication systems~\cite{du2021robust}. Although there are pilot-based and blind channel estimation methods, the perfect CSI cannot be obtained due to the estimation error, which may lead to performance deterioration. Statistical CSI, including RSRP (Reference Signal Receiving Power), RSRQ (Reference Signal Receiving Quality), RSSI (Received Signal Strength Indicator), et al., might be helpful under this condition. We would like to further investigate the joint US-BF problem in the future work.}}
\begin{subequations}\label{Eq.(05)}
\begin{align}
&\max_{\{p_{k},\mathbf{w}_{k}\}}|\mathcal{S}|,\label{Eq.(05a)}\\
\mathrm{s.t.}~&r_{k}\leq R(\overrightarrow{\gamma}_{k}),~\left\|\mathbf{w}_{k}\right\|_{2}=1,\forall{k}\in\mathcal{S},\label{Eq.(05b)}\\
&\sum\limits_{k\in\mathcal{S}}p_{k}\leq{P},\textred{~p_{k}\geq{0},\forall{k}\in\mathcal{S},}\label{Eq.(05c)}
\end{align}
\end{subequations}
where $\left|\mathcal{S}\right|$ is the cardinality of set $\mathcal{S}$, $r_{k}$ is the per-user minimum rate requirement, and $P$ denotes the power budget of the BS. Problem~\eqref{Eq.(05)} is a mixed-integer continuous-variable programming problem that involves a discrete objective function and two continuous-variable constraints about power and unit-norm BF vectors. It is difficult to obtain the global optimal solution of problem~\eqref{Eq.(05)}, even the near-optimal solution. Although the greedy heuristic search based US-BF (G-USBF) algorithm in Appendix A could be considered as a possible effective solution, it brings extremely high computational complexity especially for large-scale WCNs. In the sequel, the SCA-based US-BF optimization algorithm and the GNN-based learning algorithm are successively proposed to solve the problem~\eqref{Eq.(05)}.

\section{Design of The SCA-USBF Algorithm}
In this section, we pay our attention on designing an effective optimization algorithm for problem~\eqref{Eq.(05)} from the perspective of successive convex approximation (SCA) optimization theory. Since problem~\eqref{Eq.(05)} is non-convex, the first thing is to transform it into a tractable form via some basic mathematical transformations. \textred{One idea is to apply the uplink-downlink duality theory~\cite{schubert2004solution} to equivalently transform the downlink problem~\eqref{Eq.(05)} into a virtual uplink dual problem~\eqref{Eq.(06)}~\footnote{\textred{Similar to formula~\eqref{Eq.(01)}, the virtual uplink input-output relationship could be expressed as $\mathbf{y}=\sum\limits_{k\in{\mathcal{S}}}\sqrt{q_{k}}\overline{\mathbf{h}}_{k}s_{k}+\mathbf{n}$, where $\mathbf{y}\in\mathbb{R}^{N\times1}$ is the virtual uplink received signal at BS, $s_k$ is the virtual uplink normalized data symbol intended for the $k$-th user, and $\mathbf{n}\in\mathbb{R}^{N\times1}$ is the additive Gaussian white noise with $\mathcal{CN}(0,\mathbf{I})$. For the virtual uplink communication systems, $\mathbf{w}_{k}$ is used as the received vector for the $k$-th user. Thus, the virtual uplink received SINR of the $k$-th user can be calculated via the received signal $\mathbf{w}_{k}^{H}\mathbf{y}$.}}, i.e.,}
\begin{subequations}\label{Eq.(06)}
\begin{align}
&\max_{\left\{q_{k},\mathbf{w}_{k}\right\}}|\mathcal{S}|,\label{Eq.(06a)}\\
\mathrm{s.t.}~&r_{k}\leq{R}(\overleftarrow{\gamma}_{k}), \left\|\mathbf{w}_{k}\right\|_{2}=1,\forall{k}\in\mathcal{S},\label{Eq.(06b)}\\
&\sum\limits_{k\in\mathcal{S}}q_{k}\leq{P},\textred{~q_{k}\geq{0},\forall{k}\in\mathcal{S},}\label{Eq.(06c)}
\end{align}
\end{subequations}
where $q_{k}$ is the virtual uplink transmit power of user $k$, and $\overleftarrow{\gamma}_{k}$ represents the corresponding virtual uplink received SINR, i.e.,
\begin{equation}\label{Eq.(07)}
\overleftarrow{\gamma}_{k}=\frac{q_{k}\left|\overline{\mathbf{h}}_{k}^{H}\mathbf{w}_{k}\right|^{2}}{\sum\limits_{l\neq{k},l\in\mathcal{S}}q_{l}\left|\overline{\mathbf{h}}_{l}^{H}\mathbf{w}_{k}\right|^{2}+1}.
\end{equation}

\textred{Note that the definition (\ref{Eq.(07)}) focuses on calculating SINRs for the scheduled user set $\mathcal{S}$, with its implicit information is that the SINRs of the unscheduled users are all zero values in theory. For convenience, we further propose a new SINR definition ${\overleftarrow{\gamma}}_{k}^{(\mathcal{K})}$ which is directly calculated based on the candidate user set $\mathcal{K}$, i.e.,}

{\color{red}\begin{equation}\label{Eq.(08)}
\overleftarrow{\gamma}_{k}^{(\mathcal{K})}=\frac{q_{k}\left|\overline{\mathbf{h}}_{k}^{H}\mathbf{w}_{k}\right|^{2}}{\sum\limits_{l\neq{k},l\in\mathcal{K}}q_{l}\left|\overline{\mathbf{h}}_{l}^{H}\mathbf{w}_{k}\right|^{2}+1}.
\end{equation}}

\textred{To clearly indicate whether a user is scheduled or not, we introduce $\kappa_{k}$ as a binary variable indicator of the user state, with $\kappa_{k}=1$ if user $k$ is scheduled and $\kappa_{k}=0$ otherwise, $k\in\mathcal{K}$. Therefore, $\kappa_{k}=1$ also means that the minimum rate constraint is met for the $k$-th user, i.e., formulas $r_{k}\le{R}(\overleftarrow{\gamma}_{k})$ and $q_{k}\geq{0}$ will hold. However, $\kappa_{k}=0$ does not mean that formulas $R\left(\overleftarrow{\gamma}_{k}\right) = 0$ and $q_{k}=0$ are always true. For instance, for a candidate user set $\mathcal{K}$ and scheduled user set $\mathcal{S},\mathcal{S}\subset\mathcal{K}$. Transmission power of the BS is not always precisely exhausted for scheduled user set $\mathcal{S}$. For user $k'$ from the rest user set $\mathcal{K}\setminus\mathcal{S}$, if the residual power could not meet the minimum transmission power requirement, then we have $\kappa_{k'}=0,0<R({\overleftarrow{\gamma}_{k'}})<r_{k'},||\mathbf{w}_{k'}||_{2} = 1$. In such circumstance, $\kappa_{k'}r_{k'}\le{R}(\overleftarrow{\gamma}_{k'})$ holds, but $\kappa_{k'}=0$, i.e., $\kappa_{k'}\notin\mathcal{S}$. Meanwhile, for user $k\in\mathcal{S}$, $\overleftarrow{\gamma}_{k}>\overleftarrow{\gamma}_{k}^{\mathcal{(K)}}>0$ and $\kappa_{k}=1$ hold. For user $k\notin\mathcal{S}$, if $k\in\mathcal{K},k\ne{k'}$, then $\overleftarrow{\gamma}_{k}=\overleftarrow{\gamma}_{k}^{\mathcal{(K)}}=0$ and $\kappa_{k}=0$ hold. Let $\bm{\kappa}=[\kappa_{1},\kappa_{2},\cdots,\kappa_{k},\cdots,\kappa_{K}]^{T}$, problem~\eqref{Eq.(06)} is approximately written as}
{\color{red}\begin{subequations}\label{Eq.(09)}
\begin{align}
&\max_{\left\{\kappa_{k},q_{k},\mathbf{w}_{k}\right\}} \left\|\bm{\kappa}\right\|_{0},\label{Eq.(09a)}\\
\mathrm{s.t.}~&\kappa_{k}\in\{0,1\},\forall{k}\in\mathcal{K},\label{Eq.(09b)}\\
&\kappa_{k}r_{k}\leq{R}(\overleftarrow{\gamma}_{k}^{\mathcal{(K)}}), \left\|\mathbf{w}_{k}\right\|_{2}=1,\forall{k}\in\mathcal{K},\label{Eq.(09c)}\\
&\sum\limits_{k\in\mathcal{K}}q_{k}\leq{P},~q_{k}\geq{0},\forall k\in\mathcal{K}.\label{Eq.(09d)}
\end{align}
\end{subequations}}
\textred{As discussed above, for user $k\in\mathcal{S}$, $\overleftarrow{\gamma}_{k}>\overleftarrow{\gamma}_{k}^{\mathcal{(K)}}>0$ holds, and for user $k\notin\mathcal{S}$, $\kappa_{k}=0$ holds. Therefore, (\ref{Eq.(09c)}) is a more strict constraint than (\ref{Eq.(06b)}), and the solution to problem (\ref{Eq.(06)}) is the upper bound of problem (\ref{Eq.(09)}).}

The goal of problem~\eqref{Eq.(09)} is to maximize the number of scheduled users under the given constraints. Further, constraints~\eqref{Eq.(09b)} and~\eqref{Eq.(09c)} can be equivalently transformed into continuous constraint type and SINR form~\cite{he2020beamforming}, respectively. Let $\widetilde{\gamma}_{k}>0$ be the minimum SINR associated with achieving the minimum achievable rate $r_k$ for the $k$-th user. \textred{Thus, problem~\eqref{Eq.(09)} can be equivalently transformed as}
\begin{subequations}\label{Eq.(10)}
\begin{align}
&\max_{\{\kappa_{k},q_{k},\mathbf{w}_{k}\}} \left\|\bm{\kappa}\right\|_{0},\label{Eq.(10a)}\\
\mathrm{s.t.}~&0\leq\kappa_{k}\leq{1},\forall{k}\in\mathcal{K},\label{Eq.(10b)}\\
&\sum\limits_{k\in\mathcal{K}}\left(\kappa_{k}-\kappa_{k}^{2}\right)\leq{0},\label{Eq.(10c)}\\
&\kappa_{k}\widetilde{\gamma}_{k}\leq\overleftarrow{\gamma}_{k}^{(\mathcal{K})}, \left\|\mathbf{w}_{k}\right\|_{2}=1,\forall{k}\in\mathcal{K},\label{Eq.(10d)}\\
&\sum\limits_{k\in\mathcal{K}}q_{k}\leq{P},~q_{k}\geq{0},\forall{k}\in\mathcal{K}.\label{Eq.(10e)}
\end{align}
\end{subequations}
Constraints~\eqref{Eq.(10b)} and~\eqref{Eq.(10c)} assure that the value of $\kappa_{k}$ equals to either one or zero, i.e., $\kappa_{k}\in\{0,1\}$, $\forall k\in\mathcal{K}$. According to~\cite[Proposition 2]{che2014joint}, the strong Lagrangian duality holds for problem~\eqref{Eq.(10)}. Introducing similar mathematical tricks on handling constraint~\eqref{Eq.(10c)}, \textred{problem~\eqref{Eq.(10)} is reformulated as follows}
\begin{subequations}\label{Eq.(11)}
\begin{align}
&\min_{\{\kappa_{k},q_{k},\mathbf{w}_{k}\}}-\sum\limits_{k\in\mathcal{K}}\kappa_{k}+g\left(\bm{\kappa}\right)-h\left(\bm{\kappa}\right),\label{Eq.(11a)}\\
\mathrm{s.t.}~&~\eqref{Eq.(10b)},~\eqref{Eq.(10d)},~\eqref{Eq.(10e)},\label{Eq.(11b)}
\end{align}
\end{subequations}
where $\lambda$ is a proper non-negative constant, and $g\left(\bm{\kappa}\right)$ and $h\left(\bm{\kappa}\right)$ are defined respectively as
\begin{subequations}\label{Eq.(12)}
\begin{align}
g\left(\bm{\kappa}\right)&\triangleq\lambda\sum\limits_{k\in\mathcal{K}}\kappa_{k}+\lambda\left(\sum\limits_{k\in\mathcal{K}}\kappa_{k}\right)^{2},\\
h\left(\bm{\kappa}\right)&\triangleq\lambda\sum\limits_{k\in\mathcal{K}}\kappa_{k}^{2}+\lambda\left(\sum\limits_{k\in\mathcal{K}}\kappa_{k}\right)^{2}.
\end{align}
\end{subequations}

Note that the optimal receiver BF vector $\mathbf{w}_k^{(\ast)}$ for maximizing the uplink SINR $\overleftarrow{\gamma}_{k}^{(\mathcal{K})}$ of the $k$-th user is the
minimum mean square error (MMSE) filter with fixed $\{q_{k}\}$, i.e.,
\begin{equation}\label{Eq.(13)}
\mathbf{w}_{k}^{(\ast)}=\frac{\left(\mathbf{I}_{N}+\sum\limits_{k\in\mathcal{K}}q_{k}\overline{\mathbf{h}}_{k}\overline{\mathbf{h}}_{k}^{H}\right)^{-1}\overline{\mathbf{h}}_{k}}
{\left\|\left(\mathbf{I}_{N}+\sum\limits_{k\in\mathcal{K}}q_{k}\overline{\mathbf{h}}_{k}\overline{\mathbf{h}}_{k}^{H}\right)^{-1}\overline{\mathbf{h}}_{k}\right\|_{2}},
\end{equation}
where $\mathbf{I}_{N}$ denotes $N$-by-$N$ identity matrix. For fixed $\{\mathbf{w}_{k}\}$, \textred{problem~\eqref{Eq.(11)} is rewritten as}
\begin{subequations}\label{Eq.(14)}
\begin{align}
&\min_{\{\kappa_{k},q_{k}\}}-\sum\limits_{k\in\mathcal{K}}\kappa_{k}+g\left(\bm{\kappa}\right)-h\left(\bm{\kappa}\right),\label{Eq.(14a)}\\
\mathrm{s.t.}~&\widetilde{\gamma}_{k}\kappa_{k}-q_{k}\left|\overline{\mathbf{h}}_{k}^{H}\mathbf{w}_{k}\right|^{2}+\varphi_{k}(\bm{\kappa},\mathbf{q})-\phi_{k}(\bm{\kappa},\mathbf{q})\leq{0},\forall{k}\in\mathcal{K},\label{Eq.(14b)}\\
&~\eqref{Eq.(10b)},~\eqref{Eq.(10e)},\label{Eq.(14c)}
\end{align}
\end{subequations}
where $\varphi_{k}\left(\bm{\kappa},\mathbf{q}\right)$ and $\phi_{k}\left(\bm{\kappa},\mathbf{q}\right)$ are defined as
\begin{subequations}\label{Eq.(15)}
\begin{align}
\varphi_{k}(\bm{\kappa},\mathbf{q})&\triangleq\frac{1}{2}\left(\widetilde{\gamma}_{k}\kappa_{k}+\sum\limits_{l\in\mathcal{K},l\neq k}q_{l}\left|\overline{\mathbf{h}}_{l}^{H}\mathbf{w}_{k}\right|^{2}\right)^{2},\\
\phi_{k}(\bm{\kappa},\mathbf{q})&\triangleq\frac{1}{2}\widetilde{\gamma}_{k}^{2}\kappa_{k}^{2}+\frac{1}{2}\left(\sum\limits_{l\in\mathcal{K},l\neq k}q_{l}\left|\overline{\mathbf{h}}_{l}^{H}\mathbf{w}_{k}\right|^{2}\right)^{2}.
\end{align}
\end{subequations}
Problem~\eqref{Eq.(14)} belongs to the class of difference of convex programming problem, since the objective function~\eqref{Eq.(14a)} and constraint~\eqref{Eq.(14b)} are the difference of two convex functions. In the sequel, we resort to the classic SCA-based methods~\cite{nguyen2015achieving}. Using the convexity of functions $h\left(\bm{\kappa}\right)$ and $\phi\left(\bm{\kappa},\mathbf{q}\right)$, we have
\begin{equation}\label{Eq.(16)}
\begin{split}
&h(\bm{\kappa})\geq\psi\left(\bm{\kappa}\right)\triangleq h(\bm{\kappa}^{(\tau)})+\sum\limits_{k\in\mathcal{K}}h'(\bm{\kappa}^{(\tau)})(\kappa_{k}-\kappa_{k}^{(\tau)}),\\
&\phi_{k}(\bm{\kappa},\mathbf{q})\geq\varrho_{k}(\bm{\kappa},\mathbf{q})\triangleq\phi_{k}(\bm{\kappa}^{(\tau)},\mathbf{q}^{(\tau)})\\
&+\widetilde{\gamma}_{k}^{2}\kappa_{k}^{(\tau)}(\kappa_{k}-\kappa_{k}^{(\tau)})+\sum\limits_{l\in\mathcal{K},l\neq{k}}\rho_{k,l}(\mathbf{q}^{(\tau)})(q_{l}-q_{l}^{(\tau)}),
\end{split}
\end{equation}
where $h'\left(\kappa_{k}\right)\triangleq2\lambda\left(\kappa_{k}+\sum\limits_{l\in\mathcal{K}}\kappa_{l}\right)$, $\rho_{k,m}\left(\mathbf{q}\right)\triangleq\left|\overline{\mathbf{h}}_{m}^{H}\mathbf{w}_{k}\right|^{2}\sum\limits_{n\in\mathcal{K},n\neq k}q_{n}\left|\overline{\mathbf{h}}_{n}^{H}\mathbf{w}_{k}\right|^{2}$, and superscript $\tau$ is the $\tau$-th iteration of the SCA-USBF algorithm presented shortly. From the aforementioned discussions, \textred{the convex approximation problem solved at the $(\tau+1)$-th iteration of the proposed algorithm is given by}
\begin{subequations}\label{Eq.(17)}
\begin{align}
&\min_{\{\kappa_{k},q_{k}\}}-\sum\limits_{k\in\mathcal{K}}\kappa_{k}+g(\bm{\kappa})-\psi(\bm{\kappa}),\label{Eq.(17a)}\\
\mathrm{s.t.}~&\widetilde{\gamma}_{k}\kappa_{k}-q_{k}\left|\overline{\mathbf{h}}_{k}^{H}\mathbf{w}_{k}\right|^{2}+\varphi_{k}(\bm{\kappa},\mathbf{q})-\varrho_{k}(\bm{\kappa},\mathbf{q})\leq{0},\forall{k}\in\mathcal{K},\label{Eq.(17b)}\\
&~\eqref{Eq.(10b)},~\eqref{Eq.(10e)}.\label{Eq.(17c)}
\end{align}
\end{subequations}

Based on the above mathematical transformation, the SCA-USBF is summarized in \textbf{Algorithm}~\ref{Alg.(1)}.  In the description of \textbf{Algorithm}~\ref{Alg.(1)}, $\delta$ denotes the maximum permissible error, and $\upsilon^{(\tau)}$ and $\zeta^{(t)}$ denote the objective value of problem~\eqref{Eq.(11)} at the $\tau$-th iteration and problem~\eqref{Eq.(17)} at the $\tau$-th iteration and the $t$-th iteration, respectively. \textred{Note that SCA-USBF is also suitable for problem scenarios based on the Shannon capacity formula, as we just need to replace the $R(\overleftarrow{\gamma}_{k})$ with $C(\overleftarrow{\gamma}_{k})$ in problems (\ref{Eq.(05)}), (\ref{Eq.(06)}), and (\ref{Eq.(09)}), and replace the minimum SINR $\widetilde{\gamma}_{k}$ in problem (10b) for achieving minimum achievable rate $r_k$ with $\widetilde{\gamma}'_{k} = 2^{r_k}-1$. }The convergence of SCA-USBF can be guaranteed by the monotonic boundary theory. \textred{To speed up the convergence of SCA-USBF, we can first filter out the users which meets constraints~\eqref{Eq.(05b)} and~\eqref{Eq.(05c)} by adopting a single user communication with the maximum ratio transmission (MRT) and full power transmission, thus, at least one user could be scheduled in such circumstance.}

\begin{algorithm}[!ht]
\caption{The SCA-USBF Algorithm for Problem~\eqref{Eq.(10)}}\label{Alg.(1)}
\begin{algorithmic}[1]
\STATE Let $t=0$, $\tau=0$, $\lambda=10^{-2}$ and $\delta=10^{-5}$. Initialize the BF vectors $\{\mathbf{w}_{k}^{(0)}\}$ and downlink power vectors $\{p_{k}^{(0)}\}$, such that constraint~\eqref{Eq.(05b)} and~\eqref{Eq.(05c)} are satisfied. \label{Alg.(1-1)}
\STATE Initialize $\zeta^{(0)}$ and $\upsilon^{(0)}$, calculate the downlink SINR $\overrightarrow{\gamma}_{k}$ via $\{p_{k}^{(0)},\mathbf{w}_{k}^{(0)}\}$ and Eq.~\eqref{Eq.(02)}, and obtain the uplink power vector $\mathbf{q}=[q_{1},\cdots,q_{K^{\ast}}]^{T}$ with
\begin{equation}\label{Eq.(18)}
\mathbf{q}=\bm{\Psi}^{-1}\mathbf{I}_{K^{\ast}},
\end{equation}
where $\mathbf{I}_{K^{\ast}}$ is the all-one vector with $K^{\ast}$ dimensions, and matrix $\bm{\Psi}$ is given by
\begin{equation}\label{Eq.(19)}
[\bm{\Psi}]_{k,l}=\left\{
\begin{aligned}
\frac{|\overline{\mathbf{h}}_{k}^{H}\mathbf{w}_{k}|^{2}}{\overrightarrow{\gamma}_{k}},k=l,\\
-|\overline{\mathbf{h}}_{l}^{H}\mathbf{w}_{k}|^{2},k\neq{l}.
\end{aligned}
\right.
\end{equation}
\STATE Let $t\leftarrow{t+1}$. Solve problem~\eqref{Eq.(17)} to obtain $\{\kappa_{k}^{(t)},q_{k}^{(t)}\}$ and $\zeta^{(t)}$.\label{Alg.(1-3)}
\STATE If $|\frac{\zeta^{(t)}-\zeta^{(t-1)}}{\zeta^{(t-1)}}|\leq\delta$, go to Step~\ref{Alg.(1-5)}. Otherwise, go to Step~\ref{Alg.(1-3)}.\label{Alg.(1-4)}
\STATE Let $\tau\leftarrow\tau+1$, update $\{\mathbf{w}_{k}^{(\tau)}\}$ with $\{q_{k}^{(t)}\}$ and Eq.~\eqref{Eq.(13)}, and obtain the objective value $\upsilon^{(\tau)}$. If $|\frac{\upsilon^{(\tau)}-\upsilon^{(\tau-1)}}{\upsilon^{(\tau-1)}}|\leq\delta$, stop iteration and go to Step~\ref{Alg.(1-6)}. Otherwise, go to Step~\ref{Alg.(1-3)}.\label{Alg.(1-5)}
\STATE Calculate the uplink SINR $\overleftarrow{\gamma}_{k}$ via $\{q_{k}^{(t)},\mathbf{w}_{k}^{(\tau)}\}$ and Eq.~\eqref{Eq.(07)}, and obtain the downlink power vector $\mathbf{p}=[p_{1},\cdots,p_{K^{\ast}}]^{T}$ with\label{Alg.(1-6)}
\begin{equation}\label{Eq.(20)}
\mathbf{p}=\mathbf{D}^{-1}\mathbf{I}_{K^{\ast}},
\end{equation}
where matrix $\mathbf{D}$ is given by
\begin{equation}\label{Eq.(21)}
[\mathbf{D}]_{k,l}=\left\{
\begin{aligned}
\frac{|\overline{\mathbf{h}}_{k}^{H}\mathbf{w}_{k}|^{2}}{\overleftarrow{\gamma}_{k}},k=l,\\
-|\overline{\mathbf{h}}_{k}^{H}\mathbf{w}_{l}|^{2},k\neq{l}.
\end{aligned}
\right.
\end{equation}
\STATE Calculate the objective function value, then output the US, PA and BF vectors $\{\kappa_{k},p_{k},\mathbf{w}_{k}\}$.
\end{algorithmic}
\end{algorithm}

\section{Design of The J-USBF Algorithm}
\textred{In this section, the transformation of problem~\eqref{Eq.(05)} to problem~\eqref{Eq.(10)} is inherited, where the BF vector has an analytical solution. We focus on proposing the J-USBF learning algorithm to output the joint US-BF strategy. Specifically, we first introduce the graph representation method of single-cell WCNs, then design the JEEPON model to learn the US-PA strategy, and combine the BF analytical solution to achieve the J-USBF learning algorithm, which is summarized as \textbf{Algorithm}~\ref{Alg.(03)}.} In the sequel, we focus on studying the design of JEEPON and the corresponding training framework.
\begin{algorithm}[!ht]
\caption{The J-USBF Learning Algorithm}\label{Alg.(03)}
\begin{algorithmic}[1]
\REQUIRE $\mathcal{D}=\{\mathbf{h}_{k}\}$: Testing sample with $K$ users; \\
~\quad $\bm{\Theta}$: The trainable parameters of JEEPON. \\
\ENSURE The optimization strategy $\{\kappa_{k}^{(\ast)},q_{k}^{(\ast)},\mathbf{w}_{k}^{(\ast)}\}$ of sample $\mathcal{D}$.
\STATE Construct graph $\mathcal{G}(\mathcal{V},\mathcal{E})$ for sample $\mathcal{D}$ via the graph representation module.
\STATE Input graph $\mathcal{G}(\mathcal{V},\mathcal{E})$ to JEEPON and obtain the US-PA strategy $\{\kappa_{k}^{(\ast)},q_{k}^{(\ast)}\}$.
\STATE Calculate the BF vectors $\{\mathbf{w}_{k}^{(\ast)}\}$ via Eq.~\eqref{Eq.(13)}, and output the strategy $\{\kappa_{k}^{(\ast)},q_{k}^{(\ast)},\mathbf{w}_{k}^{(\ast)}\}$.
\STATE Calculate the uplink SINR $\overleftarrow{\gamma}_{k}$ via $\{q_{k}^{(\ast)},\mathbf{w}_{k}^{(\ast)}\}$ and Eq.~\eqref{Eq.(07)}, and obtain the downlink power vector $\{p_{k}^{(\ast)}\}$ via Eq.~\eqref{Eq.(21)}.
\end{algorithmic}
\end{algorithm}

\subsection{Problem Transformation and Loss Function Definition}
Different from the proposed SCA-USBF that alternately updates the BF vectors, in the sequel, it is regarded as intermediate variables about the virtual uplink power vectors. Taking~\eqref{Eq.(13)} into~\eqref{Eq.(08)}, the uplink received SINR of user $k$ is rewritten as
\begin{equation}\label{Eq.(22)}
\widehat{\gamma}_{k}=\frac{q_{k}|\overline{\mathbf{h}}_{k}^{H}\bm{\Lambda}^{-1}\overline{\mathbf{h}}_{k}|^{2}}{\sum\limits_{l\neq{k},l\in\mathcal{K}}q_{l}|\overline{\mathbf{h}}_{l}^{H}\bm{\Lambda}^{-1}\overline{\mathbf{h}}_{k}|^{2}+|\bm{\Lambda}^{-1}\overline{\mathbf{h}}_{k}|_{2}^{2}},
\end{equation}
where $\bm{\Lambda}=\mathbf{I}_{N}+\sum\limits_{k\in\mathcal{K}}q_{k}\overline{\mathbf{h}}_{k}\overline{\mathbf{h}}_{k}^{H}$. Thus, problem~\eqref{Eq.(10)} is formulated as follows
\begin{subequations}\label{Eq.(23)}
\begin{align}
&\max_{\{\kappa_{k},q_{k}\}} \sum\limits_{k\in\mathcal{K}}\kappa_{k},\label{Eq.(23a)}\\
\mathrm{s.t.}~&{0}\leq\kappa_{k}\leq{1},\forall{k}\in\mathcal{K},\label{Eq.(23b)}\\
&\sum\limits_{k\in\mathcal{K}}(\kappa_{k}-\kappa_{k}^{2})\leq{0},\label{Eq.(23c)}\\
&\kappa_{k}\widetilde{\gamma}_{k}\leq\widehat{\gamma}_{k},\forall{k}\in\mathcal{K},\label{Eq.(23d)}\\
&\sum\limits_{k\in\mathcal{K}}q_{k}\leq{P},~q_{k}\geq{0},\forall{k}\in\mathcal{K}.\label{Eq.(23e)}
\end{align}
\end{subequations}

\textred{To facilitate the design of JEEPON, incorporating partially the constraints into the objective function, the violation-based Lagrangian relaxation method~\cite{fioretto2020predicting} is adopted to formulate problem~\eqref{Eq.(23)} as an unconstrained optimization problem. Observe the constraints constraints~\eqref{Eq.(23b)} and~\eqref{Eq.(23e)} only contain single optimization variables that can be satisfied by subsequent projection-based methods. For constraints~\eqref{Eq.(23c)} and~\eqref{Eq.(23d)}, we introduce the non-negative Lagrangian multipliers $\{\mu,\nu\in\mathbb{R}^{+}\}$ to capture how much the constraints are violated.} Thus, the partial Lagrangian relaxation function of problem~\eqref{Eq.(23)} is given by
\begin{equation}\label{Eq.(24)}
\begin{aligned}
\mathcal{L}(\bm{\kappa},\mathbf{q},\mu,\nu)=-\sum\limits_{k\in\mathcal{K}}\kappa_{k}+\mu\sum\limits_{k\in\mathcal{K}}\chi_{c}^{\geq}(\kappa_{k}-\kappa_{k}^{2})+\nu\sum\limits_{k\in\mathcal{K}}\chi_{c}^{\geq}(\kappa_{k}\widetilde{\gamma}_{k}-\widehat{\gamma}_{k}),
\end{aligned}
\end{equation}
where $\chi_{c}^{\geq}(x)=\max\{x,0\}$ is the violation degree function. Further, the Lagrangian dual problem of~\eqref{Eq.(23)} is formulated as
\begin{equation}\label{Eq.(25)}
\begin{aligned}
\max_{\{\mu,\nu\}}\,\min_{\{\kappa_{k},q_{k}\}}\mathcal{L}(\bm{\kappa},\mathbf{q},\mu,\nu).
\end{aligned}
\end{equation}
\textred{To update the trainable parameters of JEEPON, a primal-dual learning framework (PDLF) is proposed to train it in an unsupervised manner, and the loss function is defined as $\mathrm{Loss}=\mathcal{L}/K$. In the sequel, we focus on describing the architecture of JEEPON and PDLF.}

\subsection{Graph Representation and Model Design}
\begin{figure}[!ht]
\centering
\includegraphics[width=0.5\columnwidth,keepaspectratio]{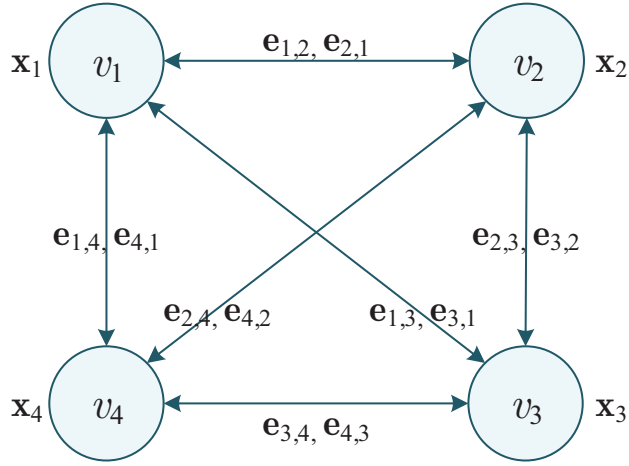}
\caption{A wireless channel graph with four UEs.}
\label{GraphStructure}
\end{figure}

\textred{WCNs can be naturally divided into undirected/directed graphs depending on the topology structures, and homogeneous/heterogeneous graphs depending on types of the communication links and user equipments (UEs)~\cite{he2021overview}. For notational convenience, a graph with node set $\mathcal{V}$ and edge set $\mathcal{E}$ is defined as $\mathcal{G}(\mathcal{V},\mathcal{E})$, where the node $v\in\mathcal{V}$ and edge $e_{u,v}\in\mathcal{E}$ (between node $u$ and node $v$) feature vectors are represented as $\mathbf{x}_{v}$ and $\mathbf{e}_{u,v}$, respectively. In the graph representation of single-cell cellular networks, we can consider the UEs as nodes and the interfering links between different UEs as edges, as shown in Fig.~\ref{GraphStructure}. In general, the node and edge features of graph $\mathcal{G}(\mathcal{V},\mathcal{E})$ mainly include CSI and other environmental information, such as user weights and Gaussian noise. In order to reduce the dimensionality of node and edge feature vectors, we consider using the orthogonality (modulo value) of CSI to represent channel gains and interferences. Therefore, the features of node $v$ and edge $e_{u,v}$ are defined as $\mathbf{x}_{v}=|\overline{\mathbf{h}}_{v}^{H}\mathbf{h}_{v}|$ and $\mathbf{e}_{u,v}=|\overline{\mathbf{h}}_{u}^{H}\mathbf{h}_{v}|$, respectively.}

\begin{figure}[!ht]
\centering
\includegraphics[width=0.9\columnwidth,keepaspectratio]{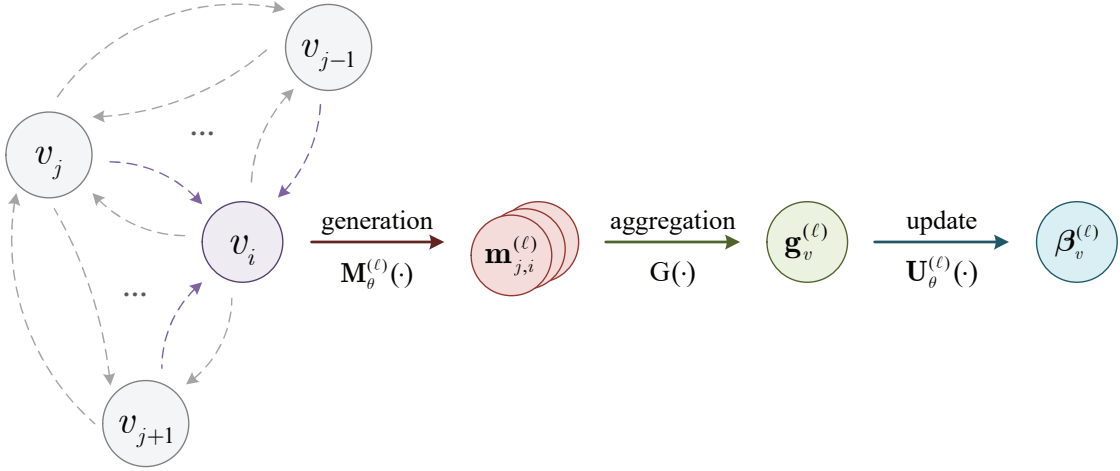}
\caption{The state update process of node $v$ at the $\ell$-th GCN layer.}
\label{MessagePassingProcess}
\end{figure}

Following the completion of the WCN graph representation, we focus on the design of JEEPON to output the US-PA strategy, where the optimization vectors are carefully defined in the representation vector of nodes. Specifically, JEEPON applies a message passing mechanism based graph convolutional network (GCN)~\cite{gilmer2017neural} to iteratively update the representation vector of node $v\in\mathcal{V}$ by aggregating features from its neighbor nodes and edges. The GCN consists of two steps, first generating and collecting messages from the first-order neighborhood nodes and edges of node $v$, and then updating the representation vector of node $v$ with the aggregated messages. After $\ell$ times of graph convolutions, the representation vector of node $v$ captures the messages within its $\ell$-hop neighborhood nodes and edges. To be specific, the update rule of the $\ell$-th GCN layer at node $v$ is formulated as
\begin{equation}\label{Eq.(26)}
\begin{aligned}
\mathbf{m}_{u,v}^{(\ell)}&=\mathbf{M}_{\theta}^{(\ell)}\left(\bm{\beta}_{u}^{(\ell-1)},\mathbf{x}_{u},\mathbf{e}_{u,v}\right),u\in\mathcal{N}_{v},\\
\mathbf{g}_{v}^{(\ell)}&=\mathbf{G}\left(F_{\mathrm{max}}\left(\{\mathbf{m}_{u,v}^{(\ell)}\}\right),F_{\mathrm{mean}}\left(\{\mathbf{m}_{u,v}^{(\ell)}\}\right)\right),v\in\mathcal{V},\\
\bm{\beta}_{v}^{(\ell)}&=\mathbf{U}_{\theta}^{(\ell)}\left(\bm{\beta}_{v}^{(\ell-1)},\mathbf{x}_{v},\mathbf{g}_{v}^{(\ell)}\right),v\in\mathcal{V},
\end{aligned}
\end{equation}
where $\mathcal{N}_{v}$ is the first-order neighborhood set of node $v$, $\bm{\beta}_{v}^{(\ell)}\triangleq[\kappa_{v},q_{v}]\in\mathbb{R}^{2}$ represents the pairwise optimization vector of node $v$ at the $\ell$-th GCN layer, and $\bm{\beta}_{v}^{(0)}$ is initialized with an all-zero vector. Therefore, when the update of the $\ell$-th graph convolution operation is completed, the representation vector of node $v$ could be formulated as $[\bm{\beta}_{v}^{(\ell)},\mathbf{x}_{v}]$. Fig.~\ref{MessagePassingProcess} illustrates the state update process of node $v$ at the $\ell$-th GCN layer. Here, $\mathrm{M}_{\theta}^{(\ell)}(\cdot)$ is a message construction function defined on each edge to generate edge message $\mathbf{m}_{u,v}^{(\ell)}\in\mathbb{R}^{m}$ by combining incoming node and edge features, where $m$ is the dimension size. $\mathbf{G}(\cdot)$ is a message aggregation function that uses the concatenation of the max function $F_{\mathrm{max}}(\cdot)$ and the mean function $F_{\mathrm{mean}}(\cdot)$ to gather the relevant edge messages $\{\mathbf{m}_{u,v}^{(\ell)}|u\in\mathcal{N}_{v}\}$ and output the aggregated message $\mathbf{g}_{v}^{(\ell)}\in\mathbb{R}^{2m}$. $\mathbf{U}_{\theta}^{(\ell)}(\cdot)$ is a state update function defined on each node, which is used to update the node representation through the aggregated message $\mathbf{g}_{v}^{(\ell)}$, node feature $\mathbf{x}_{v}$ and optimization vector $\bm{\beta}_{v}^{(\ell-1)}$. In JEEPON, function $\mathrm{M}_{\theta}^{(\ell)}(\cdot)$ and function $\mathbf{U}_{\theta}^{(\ell)}(\cdot)$ are parameterized by different neural network modules.

\begin{figure}[!ht]
\centering
\includegraphics[width=0.8\columnwidth,keepaspectratio]{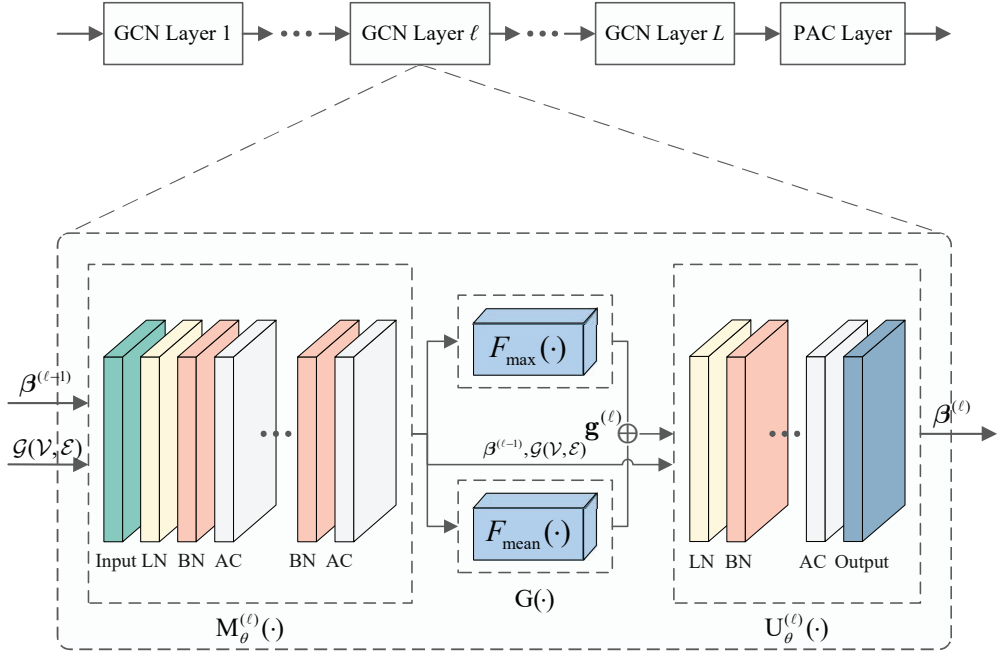}
\caption{The architecture of JEEPON.}
\label{JEEPONModel}
\end{figure}

Through the combination of several GCN layers, JEEPON can gather multi-hop node and edge features. An illustration of JEEPON is given in Fig.~\ref{JEEPONModel}, which consists of $N_{\mathrm{L}}$ GCN layers and one projection activation (PAC) layer. Each GCN layer includes an input layer, an output layer, and two different MLPs which are composed of linear (LN) layers, batch normalization (BN) layers and activation (AC) layers. In the final part of JEEPON, we utilize the PAC layer to put $\{\kappa_{k}^{(N_{\mathrm{L}})},q_{k}^{(N_{\mathrm{L}})}\}$ into the feasible region $\bm{\Omega}$, i.e.,
\begin{equation}\label{Eq.(27)}
\begin{aligned}
\bm{\Omega}\triangleq\{\bm{\kappa},\mathbf{q}:{0}\leq\kappa_{k}\leq{1},\sum\limits_{k\in\mathcal{K}}q_{k}\leq{P},q_{k}\geq{0},\forall{k}\in\mathcal{K}\}.
\end{aligned}
\end{equation}
To this end, the projection functions of the PAC layer are defined as
\begin{equation}\label{Eq.(28)}
\begin{aligned}
\kappa_{k}^{(\ast)}&=F_{\mathrm{relu}}(\kappa_{k},1),~q_{k}^{'}=F_{\mathrm{relu}}(q_{k},P),\forall{k}\in\mathcal{K}, \\
q_{k}^{(\ast)}&=\frac{P}{\mathrm{max}\{P,\sum\limits_{k\in\mathcal{K}}q_{k}^{'}\}}q_{k}^{'},\forall{k}\in\mathcal{K},
\end{aligned}
\end{equation}
where function $F_{\mathrm{relu}}(\mathbf{z},\mathbf{b})=\max\{\min\{\mathbf{z},\mathbf{0}\},\mathbf{b}\}$, and $\mathbf{b}$ is the upper bound of the input. Furthermore, due to the matrix inversion operation in the uplink SINR equation~\eqref{Eq.(22)}, it leads to a high computational overhead. To speed up the computation, the following \textit{Lemma~\ref{lemma01}} is applied to replace the direct matrix inversion by $K$ matrix iterations. Specifically, it reduces the computational complexity to $\mathcal{O}(KN^{2})$ instead of matrix inversion complexity $\mathcal{O}(KN^{2}+N^{3})$, where $\mathcal{O}(\cdot)$ is the big-$\mathcal{O}$ notation for describing the computational complexity.

\begin{lemma}\label{lemma01}
According to the Sherman-Morrison formula~\cite{sherman1950adjustment}, for an invertible square matrix $\mathbf{A}\in\mathbb{C}^{N\times{N}}$, if there exists two column vectors $\mathbf{u},\mathbf{v}\in\mathbb{C}^{N\times1}$, $1+\mathbf{v}^{H}\mathbf{A}^{-1}\mathbf{u}\neq0$, then the inverse of $\mathbf{A}$ is given by
\begin{equation}\label{Eq.(29)}
(\mathbf{A}+\mathbf{u}\mathbf{v}^{H})^{-1}=\mathbf{A}^{-1}-\frac{\mathbf{A}^{-1}\mathbf{u}\mathbf{v}^{H}\mathbf{A}^{-1}}{1+\mathbf{v}^{H}\mathbf{A}^{-1}\mathbf{u}}.
\end{equation}
Based on this formula, let $\mathbf{T}_{n}=\bm{\Lambda}^{-1},n\in\{0,1,\cdots,K\}$, then $\mathbf{T}_{n}$ can be converted into an iterative matrix product form, which is formulated as follows
\begin{equation}\label{Eq.(30)}
\mathbf{T}_{n}=
\left\{
\begin{aligned}
\mathbf{I}_{N}&, n=0,\\
\mathbf{T}_{n-1}-\frac{\mathbf{T}_{n-1}q_{n}\overline{\mathbf{h}}_{n}\overline{\mathbf{h}}_{n}^{H}\mathbf{T}_{n-1}}{1+q_{n}\overline{\mathbf{h}}_{n}^{H}\mathbf{T}_{n-1}\overline{\mathbf{h}}_{n}}&, n>0.
\end{aligned}
\right.
\end{equation}
\end{lemma}

\subsection{Primal-Dual Learning Framework}
\begin{figure}[!ht]
\centering
\includegraphics[width=0.9\columnwidth,keepaspectratio]{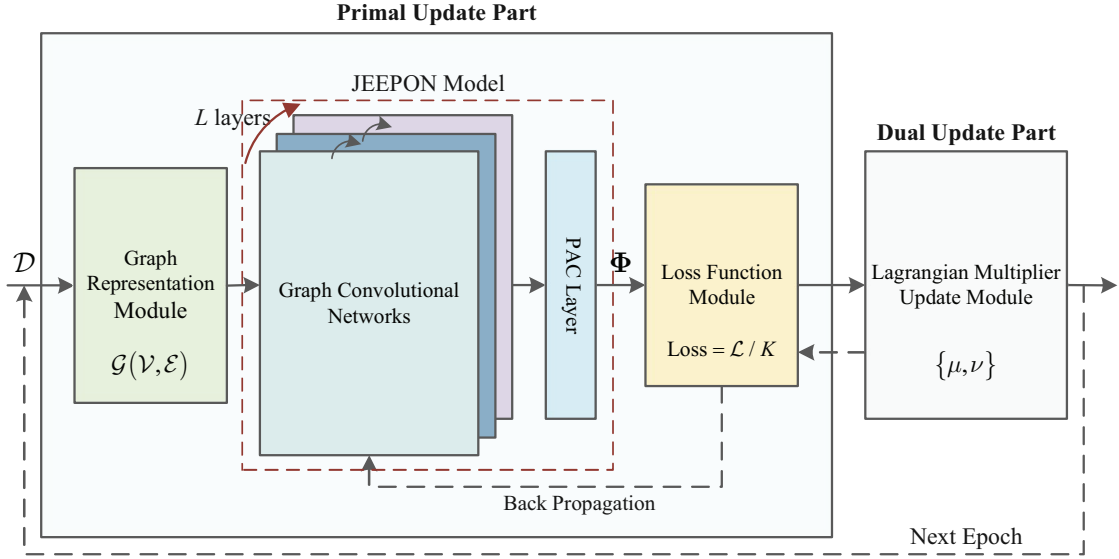}
\captionsetup{labelfont={footnotesize,color=red},font={footnotesize,color=red}}
\caption{The architecture of the PDLF.}
\label{PDLF}
\end{figure}

With regard to the aforementioned aspects, PDLF is developed for training the JEEPON model to solve the Lagrangian dual problem~\eqref{Eq.(25)}, which is composed of two parts, the primal update part and the dual update part, as shown in Fig.~\ref{PDLF}. \textred{The primal update part takes the user's historical channel data sample $\mathcal{D}\triangleq\{\mathbf{h}_{k}\}$ as input, and outputs the related US-PA strategy $\bm{\Phi}(\mathcal{D},\bm{\Theta})\triangleq\{\kappa_{k},q_{k}\}$, where $\bm{\Theta}$ is the trainable parameters of JEEPON. Specifically, it includes a graph representation module for WCN topology construction, a JEEPON model for US-PA optimization, and a loss function module for updating $\bm{\Theta}$. In the dual part, the Lagrangian multipliers $\{\mu,\nu\}$ are updated by the sub-gradient optimization method. PDLF runs two parts alternately, the former minimizes function $\mathcal{L}$ with fixed $\{\mu,\nu\}$ by updating $\bm{\Theta}$ to obtain a suitable $\{\kappa_{k},q_{k},\mathbf{w}_{k}\}$, and the latter maximizes function $\mathcal{L}$ with fixed $\{\kappa_{k},q_{k},\mathbf{w}_{k}\}$ by updating $\{\mu,\nu\}$. Therefore, the update rule of the Lagrangian multipliers $\{\mu,\nu\}$ at the $\tau$-th epoch is}
\begin{equation}\label{Eq.(31)}
\begin{aligned}
\mu^{(\tau+1)}&=\mu^{(\tau)}+\varepsilon_{\mu}\sum\limits_{k\in\mathcal{K}}\chi_{c}^{\geq}\left(\kappa_{k}^{(\tau)}-(\kappa_{k}^{(\tau)})^{2}\right), \\
\nu^{(\tau+1)}&=\nu^{(\tau)}+\varepsilon_{\nu}\sum\limits_{k\in\mathcal{K}}\chi_{c}^{\geq}\left(\kappa_{k}^{(\tau)}\widetilde{\gamma}_{k}-\widehat{\gamma}_{k}\right),
\end{aligned}
\end{equation}
where $\varepsilon_{\mu}$ and $\varepsilon_{\nu}$ is the update step-size of $\mu$ and $\nu$, respectively. In addition, the Lagrangian multipliers are updated every epoch based on the violation degree of the training datasets.
For the inner optimization of problem~\eqref{Eq.(25)}, JEEPON is proposed to transform it into a statistical learning problem, which aims to obtain appropriate optimization vectors $\{\kappa_{k},q_{k}\}$ by updating the trainable parameters of JEEPON.

\begin{algorithm}[!ht]
\caption{PDLF for Training JEEPON.}\label{Alg.(02)}
\begin{algorithmic}[1]
\REQUIRE $N_{\mathrm{e}}$: Number of epochs; \\
~\quad $\bm{\Theta}$: The trainable parameters of JEEPON; \\
~\quad $\varepsilon_{\mu},\varepsilon_{\nu}$: Step size of Lagrangian multipliers; \\
~\quad $\mathcal{D}\triangleq\{\mathcal{D}_{i}\}_{i=1}^{N_{\mathrm{ta}}}$: Training dataset with $N_{\mathrm{ta}}$ samples. \\
\ENSURE The trained JEEPON model.
\STATE Initialize the trainable parameters $\bm{\Theta}$ and the Lagrangian multipliers $\{\mu^{(0)},\nu^{(0)}\}$.
\FOR {epoch $\tau\leftarrow1,2,\cdots,N_{\mathrm{e}}$}
\STATE Initialize dual gradient variables $\{\nabla_{\mu}^{(0)},\nabla_{\nu}^{(0)}\}$.
\FOR {each sample $\mathcal{D}_{i}:i\leftarrow1,2,\cdots,N_{\mathrm{ta}}$}
\STATE Construct the graph $\mathcal{G}_{i}(\mathcal{V},\mathcal{E})$ for sample $\mathcal{D}_{i}$.
\STATE Obtain the US-PA strategy $\{\kappa_{k}^{(i)},q_{k}^{(i)}\}$ via JEEPON, and then update $\bm{\Theta}$ via the loss function module.
\STATE Update dual gradient variables:
\STATE\quad
$\nabla_{\mu}^{(i)}\leftarrow\nabla_{\mu}^{(i-1)}+\sum\limits_{k\in\mathcal{D}_{i}}\chi_{c}^{\geq}(\kappa_{k}^{(i)}-(\kappa_{k}^{(i)})^{2})$,
$\nabla_{\nu}^{(i)}\leftarrow\nabla_{\nu}^{(i-1)}+\sum\limits_{k\in\mathcal{D}_{i}}\chi_{c}^{\geq}(\kappa_{k}^{(i)}\widetilde{\gamma}_{k}-\widehat{\gamma}_{k})$.
\ENDFOR
\STATE Obtain the Lagrangian multipliers $\{\mu^{(\tau)},\nu^{(\tau)}\}$ via Eq.~\eqref{Eq.(31)}.
\ENDFOR
\end{algorithmic}
\end{algorithm}

PDLF is designed for training JEEPON. Unlike the penalty-based supervised training method in~\cite{xia2020deep}, the proposed PDLF alternately updates $\bm{\Theta}$ and $\{\mu,\nu\}$ in an unsupervised manner, as summarized in \textbf{Algorithm}~\ref{Alg.(02)}. Specifically, we generate a training dataset $\mathcal{D}\triangleq\{\mathcal{D}_{i}\}_{i=1}^{N_{\mathrm{ta}}}$ with $N_{\mathrm{ta}}$ samples, and each sample with the same size. The training stage of PDLF lasts for $N_{\mathrm{e}}$ epochs in total. \textred{In the primal update part, PDLF first constructs the graph representation for sample $\mathcal{D}_{i}$ (Setp 5), and takes it as the input of JEEPON. Then, JEEPON outputs the US-PA strategy $\bm{\Phi}(\mathcal{D}_{i},\bm{\Theta})\triangleq\{\kappa_{k}^{(i)},q_{k}^{(i)}\}$ of sample $\mathcal{D}_{i}$ (Step 6), and adopt the loss function module to update $\bm{\Theta}$ (Step 7). The sub-gradient values of $\{\mu,\nu\}$ are also stored to avoid repeated traversal of the training dataset (Steps 8-10).} Therefore, in the dual update part, $\{\mu,\nu\}$ are updated by the recorded dual gradient variables $\{\nabla_{\mu}^{(N_{\mathrm{ta}})},\nabla_{\nu}^{(N_{\mathrm{ta}})}\}$ and equation~\eqref{Eq.(31)} (Step 13).

\section{Numerical Results}
In this section, numerical results are presented for the joint US-BF optimization problem in the multiuser MISO downlink system. We first introduce the experimental environment and system parameters. Next, the convergence of SCA-USBF and J-USBF is evaluated. Then, the performance of G-USBF, SCA-USBF, and J-USBF is discussed in different system scenarios, as well as the generalizability of J-USBF. In addition, the performance of J-USBF and the convolutional neural network based US-BF (CNN-USBF) algorithm (see Appendix B) are also compared. Finally, the computational complexity of the algorithms is presented and discussed, which clearly validates the speed advantage of J-USBF.

\subsection{Experimental Setup}
\begin{figure}[ht]
\centering
\includegraphics[width=0.8\columnwidth,keepaspectratio]{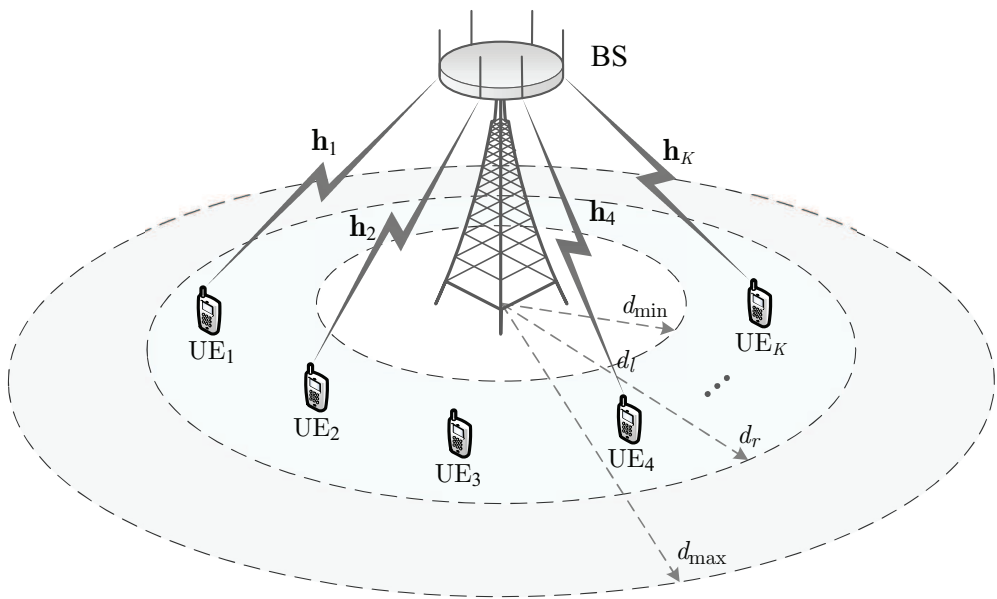} 
\caption{User distribution of the multiuser MISO downlink system.}
\label{SystemModel}
\end{figure}

In the experiment\footnote{\textred{Offline training for J-USBF is necessary and important, where numerous data is required. However, real data is quite difficult to obtain although some researchers are committed to solving this problem \cite{huang2021true-data,coronado20195G-EmPOWER,munoz2016the}, so we could only apply simulated data instead.}}, the $K$ single-antenna users are randomly distributed in the range of $(d_{l},d_{r})$ from the BS, $d_{l},d_{r}\in(d_{\mathrm{min}},d_{\mathrm{max}})$, where $d_{\mathrm{min}}=50\mathrm{m}$ is the reference distance and $d_{\mathrm{max}}=200\mathrm{m}$ denotes the cell radius, as shown in Fig.~\ref{SystemModel}. The channel of user $k$ is modeled as $\mathbf{h}_{k}=\sqrt{\rho_{k}}\widetilde{\mathbf{h}}_{k}\in\mathbb{C}^{N\times1}$ where $\widetilde{\mathbf{h}}_{k}\sim\mathcal{CN}(\mathbf{0},\mathbf{I}_{N})$ is the small-scale fading, $\varrho=3$ is the path-loss exponent, and $\rho_{k}=1/(1+(d_{k}/d_{\mathrm{min}})^{\varrho})$ denotes the long-term path-loss between user $k$ and the BS with $d_{k}$ representing the distance in meters (m). For simplicity, we assume that all users have the same additive noise variance, i.e., $\sigma_{k}^{2}=\sigma^{2},\forall{k}\in\mathcal{K}$, thus, the signal-to-noise ratio (SNR) is defined as $\mathrm{SNR}=10\log_{10}(P/\sigma^{2})$ in dB.

In the neural network module, J-USBF is implemented by $N_{\mathrm{L}}=2$ GCN layers via \emph{Pytorch}, and the functions $\mathrm{M}_{\theta}(\cdot)$ and $\mathbf{U}_{\theta}(\cdot)$ in each GCN layer are parameterized by MLPs with sizes $\mathcal{H}_{1}$ and $\mathcal{H}_{2}$, respectively. Training and test stages for J-USBF are sequentially implemented. The learing rate of J-USBF and Lagrangian multipliers are set to $\eta=5\times10^{-5}$ and $\varepsilon_{\mu},\varepsilon_{\nu}=10^{-5}$, respectively. For each configuration, we respectively prepare $N_{\mathrm{ta}}=5000$ training samples and $N_{\mathrm{te}}=500$ testing samples, where the validation split is set to $0.2$ and the training samples are randomly shuffled at each epoch. The entire training stage lasts for $N_{\mathrm{e}}=200$ epochs. According to the conclusion in~\cite[Corollary 1]{he2020beamforming}, the user minimum achievable SINR $\widetilde{\gamma}$ will be set by the system parameters $\{D,n,\epsilon\}$. Note that unless mentioned otherwise, the experiments adopt the default system parameters in Table~\ref{Tab-01}.


\begin{table}[!ht]
\centering
\renewcommand{\arraystretch}{1.1}
\caption{Default system parameters.}\label{Tab-01}
\begin{tabular}{|c|c|}
\hline
Parameters & Values \\\hline
Range of SNR & $10~\mathrm{dB}$ \\\hline
Blocklength $n$ & $128$ \\\hline
Decoding error probability $\epsilon$ & $10^{-6}$ \\\hline
Transmission data bits $D$ & $256~\mathrm{bits}$ \\\hline
BS antenna number $N$ & $32$ \\\hline
Number of candidate users $K$ & $30$ \\\hline
Maximum permissible error $\delta$ & $10^{-5}$ \\\hline
Sizes of MLPs in $\mathrm{M}_{\theta}(\cdot)$ & $\mathcal{H}_{1}=\{4,256,128,64,32,16,m\}$ \\\hline
Sizes of MLPs in $\mathbf{U}_{\theta}(\cdot)$ & $\mathcal{H}_{2}=\{3+2m,256,128,64,32,16,3\}$\\\hline
\end{tabular}
\end{table}

\subsection{Convergence Analysis of SCA-USBF and J-USBF}
\begin{figure}[ht]
\centering
{\color{red}
\begin{minipage}{0.45\linewidth}
\centering
\subfigure[SCA-USBF]{
\label{Fig-Target1}
\includegraphics[width=1.0\linewidth]{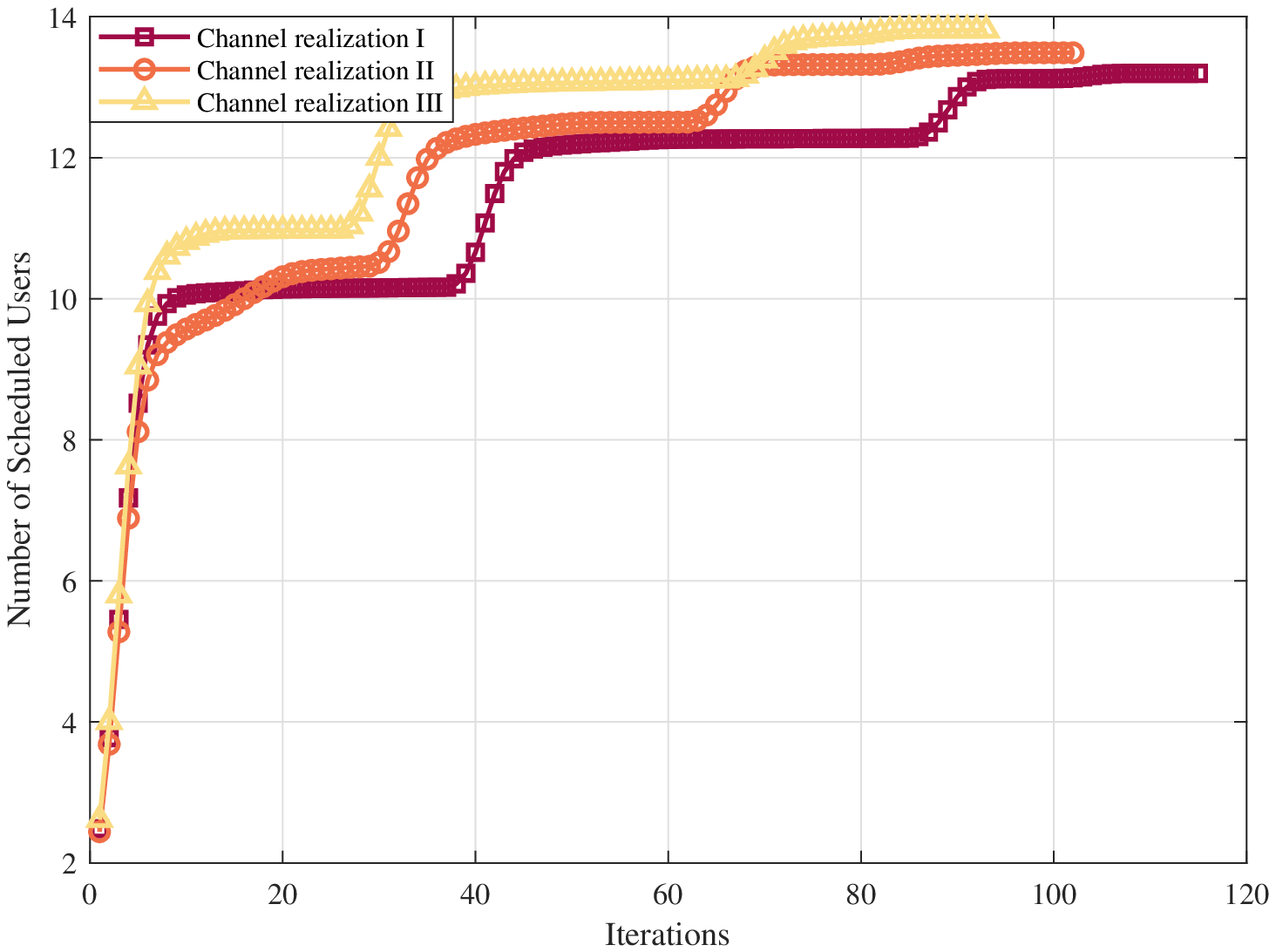}} 
\end{minipage}
\begin{minipage}{0.45\linewidth}
\centering
\subfigure[J-USBF]{
\label{Fig-Target2}
\includegraphics[width=1.0\linewidth]{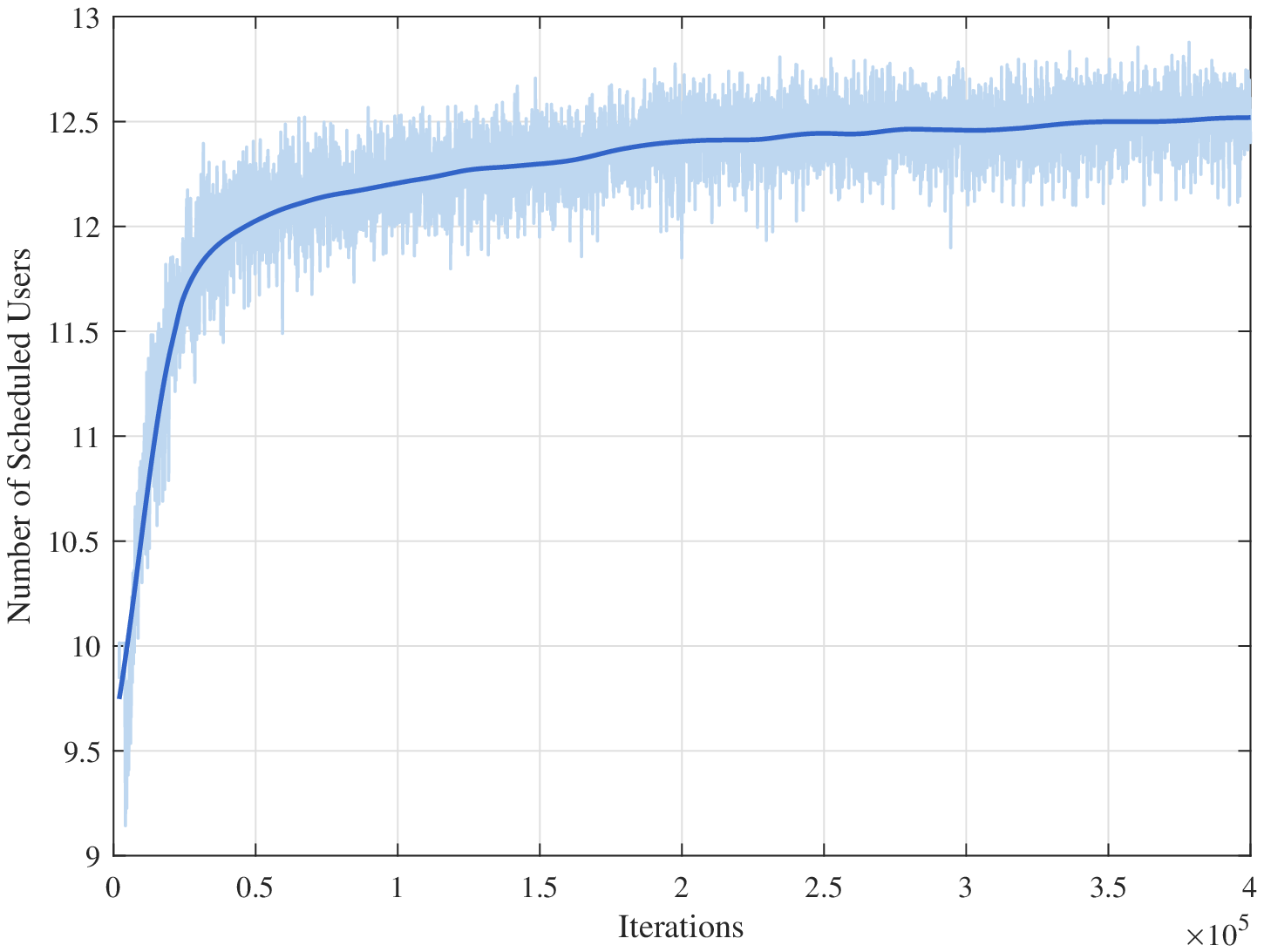}} 
\end{minipage}
}
\captionsetup{labelfont={footnotesize,color=red},font={footnotesize,color=red}}
\caption{The objective value curves of SCA-USBF and J-USBF.}
\label{EXP01_FIG}
\end{figure}

\textred{The convergence of SCA-USBF and J-USBF is evaluated in this section, where part of the system parameters are set to $K=50$ and $(d_{l},d_{r})=(60\mathrm{m},100\mathrm{m})$. Fig.~\ref{Fig-Target1} illustrates the objective value curve of SCA-USBF for different random channel realizations, indicating that SCA-USBF can reach the convergence state through iterations. Fig.~\ref{Fig-Target2} illustrates the objective value curve of J-USBF during the training stage, where the objective value of the training samples varies in the range (light blue line), and the average objective value curve (blue line) converges as the number of iterations increases to $3.5\times10^{5}$. During the testing stage, the constraint violation ratios of J-USBF for different testing samples are shown in Table~\ref{Tab-02}. It is observed that the percentage of illegal results is $2.268\%$, which is much lower than the results satisfying the constraints.} Note that the value of $\kappa_{k},\forall{k}\in\mathcal{K}$ will be set to $1$ if $0<\kappa_{k}<1$ is obtained, and all the scheduled users are filtered again with the per-user minimum SINR requirement.

\begin{table}[!ht]
\centering
\renewcommand{\arraystretch}{1.1}
\captionsetup{labelfont={color=red},font={color=red}}
\caption{Different constraint situations of J-USBF.}\label{Tab-02}
{\color{red}
\begin{tabular}{|c|c|}
\hline
Different constraint situations & Percentage of total samples \\\hline
$\kappa_{k}=0,q_{k}\geq{0},\forall{k}\in\mathcal{K}$ & $75.436\%$ \\\hline
$0<\kappa_{k}<1,q_{k}\geq{0},\forall{k}\in\mathcal{K}$ & $2.264\%$ \\\hline
$\kappa_{k}=1,\widetilde{\gamma}_{k}>\widehat{\gamma}_{k},\forall{k}\in\mathcal{K}$ & $0.004\%$ \\\hline
$\kappa_{k}=1,\widetilde{\gamma}_{k}\leq\widehat{\gamma}_{k},\forall{k}\in\mathcal{K}$ & $22.296\%$ \\\hline
\end{tabular}
}
\end{table}

\subsection{Performance and Generalizability Evaluation}
In this subsection, the performance of J-USBF, SCA-USBF and G-USBF with different system parameters are evaluated and compared. For intuitive comparison, the obtained results of SCA-USBF and J-USBF are normalized by the results of G-USBF, defined as $R_{1}=\frac{N_{\mathrm{S}}}{N_{\mathrm{G}}}\times100\%$ and $R_{2}=\frac{N_{\mathrm{J}}}{N_{\mathrm{G}}}\times100\%$, where $N_{\mathrm{S}}$, $N_{\mathrm{J}}$ and $N_{\mathrm{G}}$ are the average number of scheduled users obtained through SCA-USBF, J-USBF and G-USBF, respectively. In addition, we also define the result percentage of CNN-USBF and J-USBF as $R_{3}=\frac{N_{\mathrm{C}}}{N_{\mathrm{J}}}\times100\%$, where $N_{\mathrm{C}}$ is the number of scheduled users obtained through CNN-USBF.

\subsubsection{Performance with Various $K$ and $(d_{l},d_{r})$}
\begin{figure}[!ht]
\centering
\includegraphics[width=0.6\columnwidth,keepaspectratio]{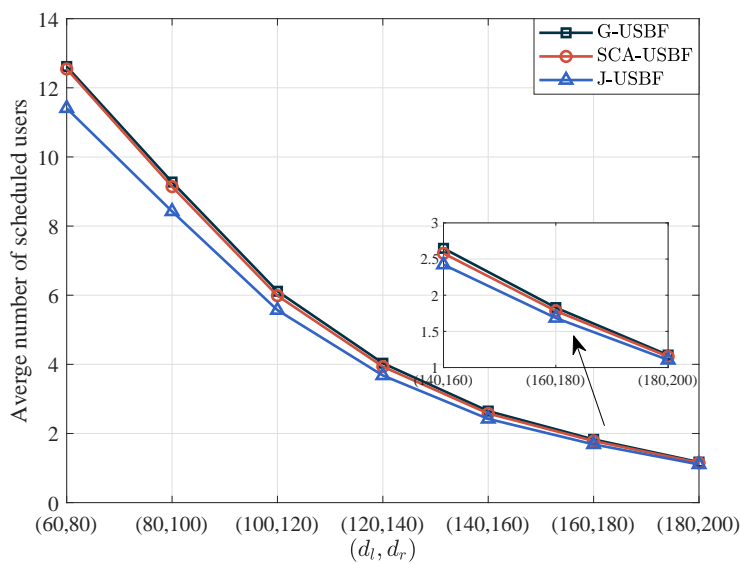} 
\caption{Average number of scheduled users with various $(d_{l},d_{r})$.}
\label{Fig-Distance}
\end{figure}

This experiment investigates the influences of $K$ and $(d_{l},d_{r})$ and compares the performance of J-USBF with G-USBF and SCA-USBF, as well as with CNN-USBF in large-scale user scenarios. Table~\ref{Tab-03} shows that when $K$ is small, the performance of J-USBF is closer to that of G-USBF, because sufficient system resources are conducive to model optimization. J-USBF remains stable when $K$ changes from 20 to 50, and there exist only $2.56\%$ performance degradation at most. Besides, the performance gain of J-USBF improves with the distance interval changes from $20\mathrm{m}$ to $40\mathrm{m}$, since the smaller distance interval leads to the lack of diversity for each user, which brings more difficulties to the learning of J-USBF. In Fig.~\ref{Fig-Distance}, we show the average performance of these three algorithms with different $(d_{l},d_{r})$. It suggests that J-USBF could achieve a more stable and closer performance compared with G-USBF as $(d_{l},d_{r})$ increases. Owing to the fact that the number of scheduled users is reduced with the increase of $(d_{l},d_{r})$, and the obtained results are more homogeneous, which is beneficial to the learning of J-USBF.

\begin{table*}[!ht]
\centering
\fontsize{8}{8}\selectfont
\renewcommand{\arraystretch}{1.5}
\newcolumntype{C}[1]{>{\centering}p{#1}}
\caption{Performance normalized by G-USBF with various $K$.}\label{Tab-03}
\begin{tabular}{|C{3em}|c|c|c|c|c|c|c|c|}
\hline
\multirow{3}{*}{$K$} & \multicolumn{8}{c|}{$R_{1}$ and $R_{2}$ with varying $(d_{l},d_{r})$} \\\cline{2-9}
& \multicolumn{2}{c|}{$(50\mathrm{m},70\mathrm{m})$} & \multicolumn{2}{c|}{$(60\mathrm{m},80\mathrm{m})$} & \multicolumn{2}{c|}{$(50\mathrm{m},90\mathrm{m})$} & \multicolumn{2}{c|}{$(60\mathrm{m},100\mathrm{m})$} \\\cline{2-9}
& $R_{1}$ & $R_{2}$  & $R_{1}$ & $R_{2}$ & $R_{1}$ & $R_{2}$ & $R_{1}$ & $R_{2}$ \\\hline
10 & $100\%$ & \blue{$95.68\%$} & $99.98\%$ & \blue{$94.70\%$} & $100\%$ & \blue{$93.94\%$} & $99.89\%$ & \blue{$93.41\%$} \\\cline{1-9}
20 & $99.67\%$ & \blue{$90.04\%$} & $99.64\%$ & \blue{$91.35\%$} & $99.52\%$  & \blue{$92.02\%$} & $99.22\%$ & \blue{$92.32\%$} \\\cline{1-9}
30 & $99.94\%$ & \blue{$89.68\%$} & $99.63\%$ & \blue{$90.33\%$} & $98.80\%$ & \blue{$90.57\%$} & $98.77\%$ & \blue{$91.25\%$} \\\cline{1-9}
40 & $99.86\%$ & \blue{$88.91\%$} & $99.54\%$ & \blue{$89.86\%$} & $98.52\%$ & \blue{$90.27\%$} & $98.24\%$ & \blue{$91.08\%$} \\\cline{1-9}
50 & $99.84\%$ & \blue{$88.15\%$} & $98.73\%$ & \blue{$88.79\%$} & $97.48\%$ & \blue{$89.46\%$} & $97.10\%$ & \blue{$90.15\%$} \\\hline
\end{tabular}
\end{table*}

\textred{Considering large-scale user scenarios, we focus on the performance comparison of J-USBF and CNN-USBF, whereas ignoring G-USBF and SCA-USBF due to the high computational overhead. Table~\ref{Tab-04} shows that the performance gap between CNN-USBF and J-USBF widens as $K$ increases, especially when $K=200$ and $(d_{l},d_{r})=(60\mathrm{m},100\mathrm{m})$, the performance of the former can only reach $87.36\%$ of the latter. This indicates that incorporating WCN topology information into model learning is helpful for performance improvement and stability maintenance.}

\begin{table}[!ht]
\centering
\fontsize{8}{8}\selectfont
\renewcommand{\arraystretch}{1.5}
\newcolumntype{C}[1]{>{\centering}p{#1}}
\captionsetup{labelfont={color=red},font={color=red}}
\caption{Performance normalized by J-USBF with various $K$.}\label{Tab-04}
{\color{red}\begin{tabular}{|C{3em}|c|c|c|c|}
\hline
\multirow{2}{*}{$K$} & \multicolumn{4}{c|}{$R_{3}$ with varying $(d_{l},d_{r})$} \\\cline{2-5}
& $(50\mathrm{m},70\mathrm{m})$ & $(60\mathrm{m},80\mathrm{m})$ & $(50\mathrm{m},90\mathrm{m})$ & $(60\mathrm{m},100\mathrm{m})$ \\\cline{1-5}
50  & $93.06\%$ & $96.26\%$ & $92.42\%$ & $93.95\%$ \\\cline{1-5}
100 & $92.89\%$ & $90.83\%$ & $91.42\%$ & $90.80\%$ \\\cline{1-5}
150 & $89.14\%$ & $90.53\%$ & $90.59\%$ & $89.46\%$ \\\cline{1-5}
200 & $88.75\%$ & $88.38\%$ & $87.51\%$ & $87.36\%$ \\\hline
\end{tabular}}
\end{table}

\subsubsection{Performance with Various SNR Settings}
\begin{figure*}[ht]
\centering
\subfigure[$(d_{l},d_{r})=(60\mathrm{m},90\mathrm{m})$]{
\label{Fig-SNR1}
\includegraphics[width=0.45\linewidth]{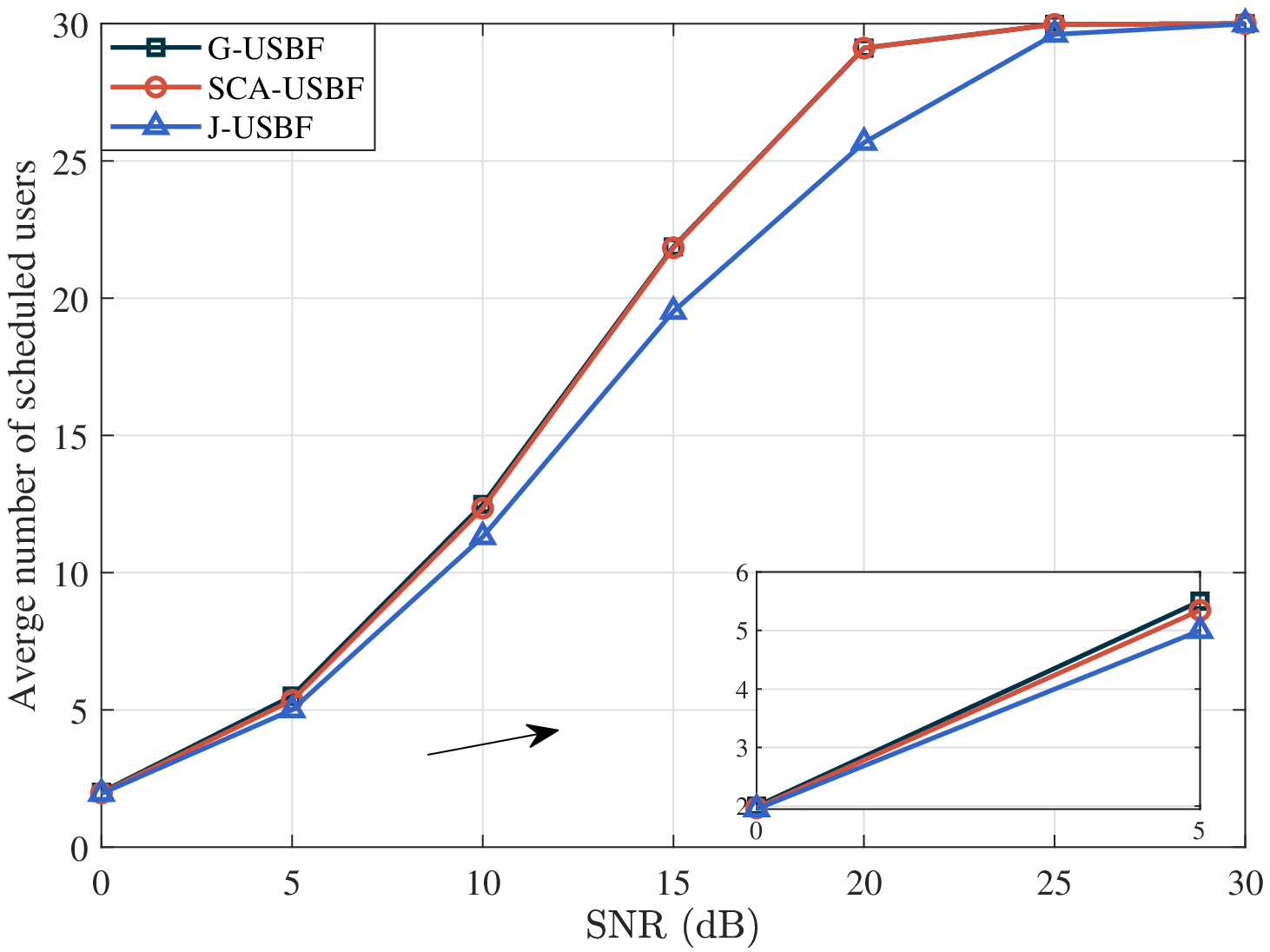}}
\subfigure[$(d_{l},d_{r})=(100\mathrm{m},150\mathrm{m})$]{
\label{Fig-SNR2}
\includegraphics[width=0.45\linewidth]{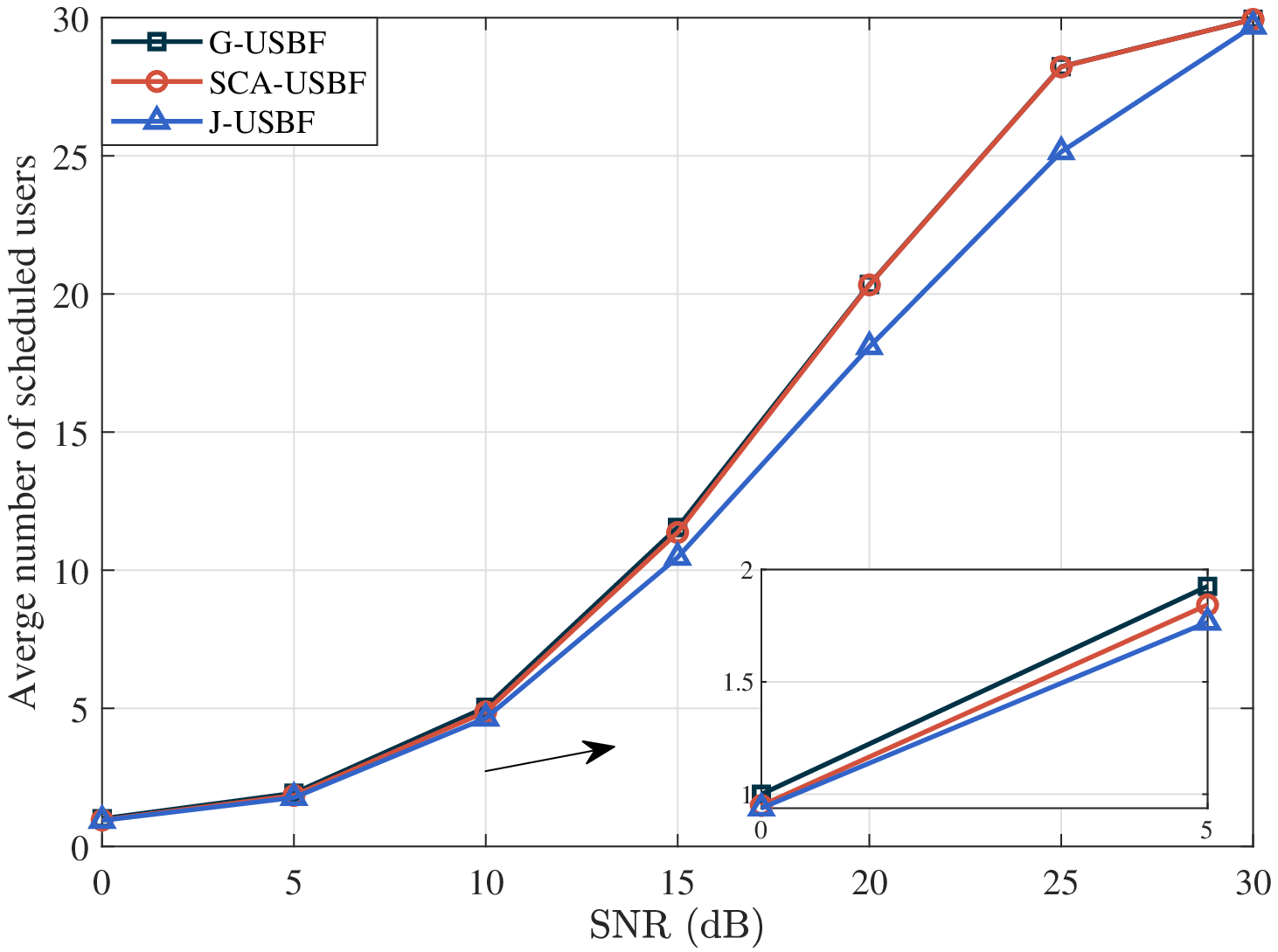}}
\caption{Performance of the algorithms with various SNR settings.}
\label{Fig-SNR}
\end{figure*}

This experiment compares the performance of J-USBF, SCA-USBF and G-USBF with different SNR settings, and the obtained results are summarized in Table~\ref{Tab-05}. It is observed that J-USBF achieves competitive performance (larger than $90.77\%$) with $\mathrm{SNR}=5$~dB, while SCA-USBF maintains over $95.73\%$ near-optimal performance compared with G-USBF. Although the performance gap of J-USBF is enlarged as $K$ increase, the trend of degradation is rather slow. For the configuration $\mathrm{SNR}=15$~dB and $(d_{l},d_{r})=(50\mathrm{m},100\mathrm{m})$, J-USBF obtains only a $1.78\%$ relative performance gap with G-USBF when $K$ changes from 20 to 50. Even when $(d_{l},d_{r})=(100\mathrm{m},150\mathrm{m})$, J-USBF still maintains a stable performance. \textred{Moreover, Fig.~\ref{Fig-SNR} illustrates the gap between the SCA-USBF and J-USBF increases while SNR changes from 0~dB to 20~dB. With the increase of SNR, channel condition becomes better and more users might meet the requirement of QoS. Therefore, the solution space for problem~\eqref{Eq.(05)} enlarges and SCA-USBF shows its advantages under this condition, because it obtains optimal/suboptimal results. On the other hand, J-USBF is difficult to capture the optimal value as the solution space increases in such circumstance. However, the gap between the SCA-USBF and J-USBF decreases when SNR changes from 20~dB to 30~dB, since much better channel condition is sufficient for serving all the users.}


\begin{table*}[!ht]
\centering
\fontsize{8}{8}\selectfont
\renewcommand{\arraystretch}{1.5}
\newcolumntype{C}[1]{>{\centering}p{#1}}
\caption{Performance normalized by G-USBF with various SNR settings.}\label{Tab-05}
\begin{tabular}{|C{4em}|C{3em}|c|c|c|c|c|c|c|c|}
\hline
\multirow{3}{*}{$\mathrm{SNR}(\mathrm{dB})$} & \multirow{3}{*}{$K$} & \multicolumn{8}{c|}{$R_{1}$ and $R_{2}$ with varying $(d_{l},d_{r})$} \\\cline{3-10}
&& \multicolumn{2}{c|}{$(60\mathrm{m},90\mathrm{m})$} & \multicolumn{2}{c|}{$(90\mathrm{m},120\mathrm{m})$} & \multicolumn{2}{c|}{$(50\mathrm{m},100\mathrm{m})$} & \multicolumn{2}{c|}{$(100\mathrm{m},150\mathrm{m})$} \\\cline{3-10}
&& $R_{1}$ & $R_{2}$  & $R_{1}$ & $R_{2}$ & $R_{1}$ & $R_{2}$ & $R_{1}$ & $R_{2}$ \\\hline
\multirow{3}{*}{5}  & 10 & $98.96\%$ & \blue{$92.13\%$} & $98.43\%$ & \blue{$94.50\%$} & $99.37\%$ & \blue{$92.56\%$} & $97.74\%$ & \blue{$93.17\%$} \\\cline{2-10}
& 20 & $98.22\%$ & \blue{$91.08\%$} & $97.82\%$ & \blue{$92.43\%$} & $98.82\%$ & \blue{$91.15\%$} & $96.59\%$ & \blue{$92.03\%$} \\\cline{2-10}
& 30 & $97.17\%$ & \blue{$90.77\%$} & $96.58\%$ & \blue{$91.04\%$} & $97.75\%$ & \blue{$90.83\%$} & $95.73\%$ & \blue{$91.66\%$} \\\hline
\multirow{3}{*}{15} & 20 & $99.99\%$ & \blue{$90.15\%$} & $99.60\%$ & \blue{$90.69\%$} & $100\%$ & \blue{$90.48\%$} & $99.17\%$ & \blue{$91.15\%$} \\\cline{2-10}
& 30 & $99.87\%$ & \blue{$89.20\%$} & $99.61\%$ & \blue{$89.78\%$} & $99.97\%$ & \blue{$89.52\%$} & $98.46\%$ & \blue{$90.60\%$} \\\cline{2-10}
& 50 & $99.64\%$ & \blue{$88.34\%$} & $99.55\%$ & \blue{$89.06\%$} & $99.80\%$ & \blue{$88.70\%$} & $97.36\%$ & \blue{$89.94\%$} \\\hline
\end{tabular}
\end{table*}

\subsubsection{Performance with Various SINR Requirements}
The ultimate scheduling results of the investigated problem are significantly affected by the per-user minimum SINR requirement, where value $\widetilde{\gamma}=F_{\mathrm{\gamma}}(D,n,\epsilon)$ is obtained with different system parameters $D$, $n$, and $\epsilon$, and the results are summarized in Table~\ref{Tab-06}. From the table, It is observed that the average performance of J-USBF remains above $88.97\%$ compared with G-USBF under different SINR requirements and user distribution distances. However, one needs to point out that with the reduction of SINR requirements, the performance improvement of J-USBF is lower than G-USBF, especially when $(d_{l},d_{r})=(60\mathrm{m},80\mathrm{m})$. Therefore, J-USBF shows a slight performance degradation compared with G-USBF when the SINR requirement is reduced, while the performance improvement of SCA-USBF increases at the same time.

\begin{table*}[!ht]
\centering
\fontsize{8}{8}\selectfont
\renewcommand{\arraystretch}{1.5}
\newcolumntype{C}[1]{>{\centering}p{#1}}
\caption{Performance normalized by G-USBF with various SINR requirements.}\label{Tab-06}
\begin{tabular}{|C{7em}|C{3em}|c|c|c|c|c|c|c|c|}
\hline
\multirow{3}{*}{$F_{\mathrm{\gamma}}(D,n,\epsilon)$} & \multirow{3}{*}{$\widetilde{\gamma}$} & \multicolumn{8}{c|}{$R_{1}$ and $R_{2}$ with varying $(d_{l},d_{r})$} \\\cline{3-10}
&& \multicolumn{2}{c|}{$(60\mathrm{m},80\mathrm{m})$} & \multicolumn{2}{c|}{$(80\mathrm{m},100\mathrm{m})$} & \multicolumn{2}{c|}{$(60\mathrm{m},100\mathrm{m})$} & \multicolumn{2}{c|}{$(80\mathrm{m},120\mathrm{m})$} \\\cline{3-10}
&& $R_{1}$ & $R_{2}$ & $R_{1}$ & $R_{2}$ & $R_{1}$ & $R_{2}$ & $R_{1}$ & $R_{2}$ \\\hline
$(256,256,10^{-6})$ & $1.633$ & $99.92\%$ & \blue{$88.97\%$} & $99.96\%$ & \blue{$90.36\%$} & $99.97\%$ & \blue{$89.82\%$} & $99.91\%$ & \blue{$91.03\%$} \\\hline
$(256,128,10^{-6})$ & $5.054$ & $99.63\%$ & \blue{$90.33\%$} & $98.30\%$ & \blue{$91.87\%$} & $98.77\%$ & \blue{$91.25\%$} & $98.36\%$ & \blue{$92.78\%$} \\\hline
$(256,96,10^{-6})$ & $9.291$ & $96.41\%$ & \blue{$90.79\%$} & $94.22\%$ & \blue{$92.94\%$} & $95.84\%$ & \blue{$91.84\%$} & $95.38\%$ & \blue{$93.62\%$} \\\hline
$(256,64,10^{-6})$ & $27.97$ & $95.58\%$ & \blue{$91.05\%$} & $93.95\%$ & \blue{$93.05\%$} & $94.55\%$ & \blue{$92.76\%$} & $94.19\%$ & \blue{$94.08\%$} \\\hline
\end{tabular}
\end{table*}

\subsubsection{Generalizability with Various User Distributions}
Generalizability is another critical evaluation property for J-USBF, and it focuses on investigating whether the trained network model has the ability to perform well in unknown WCN scenarios. To test the generalizability, J-USBF is trained from a certain scenario whose system parameters are different from the test ones. Specifically, J-USBF is trained with $(d_{l},d_{r})=(100\mathrm{m},120\mathrm{m})$, then the trained model is applied to the test scenarios with different $(d_{l},d_{r})$, without any further training\footnote{\textred{For scenarios with different number of users $K$, number of antennas $N$ and SINR requirements $\widetilde{\gamma}$, the generalizability of J-USBF performs poorly and needs to be further optimized.}}. Table~\ref{Tab-07} shows comparison results of G-USBF and J-USBF, where $R_{4}=\frac{N_{\mathrm{J},(100,120)}}{N_{\mathrm{G}}}\times100\%$ represents the average performance of J-USBF normalized by G-USBF and $N_{\mathrm{J},(100,120)}$ is the average number of scheduled users using J-USBF. Form the table, it is observed that J-USBF performs well over the neighboring user distribution distances when the test distance interval is 40m. Moreover, when $(d_{l},d_{r})=(80\mathrm{m},100\mathrm{m})$ and there is no intersection with $(100\mathrm{m},120\mathrm{m})$, the performance of J-USBF is still acceptable at $K=10$. Based on the aforementioned analysis, our proposed J-USBF can be well generalized to scenarios with neighboring user distribution distances.

\begin{table}[!ht]
\centering
\fontsize{8}{8}\selectfont
\renewcommand{\arraystretch}{1.5}
\newcolumntype{C}[1]{>{\centering}p{#1}}
\caption{Generalizability with various user distributions.}\label{Tab-07}
\begin{tabular}{|C{3em}|c|c|c|c|c|c|c|c|}
\hline
\multirow{3}{*}{$K$} & \multicolumn{8}{c|}{$N_{\mathrm{G}}$ and $R_{4}$ with varying $(d_{l},d_{r})$} \\\cline{2-9}
& \multicolumn{2}{c|}{$(100\mathrm{m},120\mathrm{m})$} &
\multicolumn{2}{c|}{$(80\mathrm{m},100\mathrm{m})$} & \multicolumn{2}{c|}{$(80\mathrm{m},120\mathrm{m})$} &
 \multicolumn{2}{c|}{$(100\mathrm{m},140\mathrm{m})$} \\\cline{2-9}
& $N_{\mathrm{G}}$ & $R_{4}$ & $N_{\mathrm{G}}$ & $R_{4}$ & $N_{\mathrm{G}}$ & $R_{4}$ & $N_{\mathrm{G}}$ & $R_{4}$ \\\hline
10 & $5.068$ & \blue{$94.97\%$} & $7.504$ & \blue{$86.06\%$} & $6.36$ & \blue{$91.86\%$} & $4.394$ & \blue{$92.13\%$} \\\hline
20 & $5.624$ & \blue{$93.92\%$} & $8.548$ & \blue{$84.63\%$} & $7.496$ & \blue{$89.58\%$} & $5.068$ & \blue{$91.46\%$} \\\hline
30 & $5.924$ & \blue{$92.69\%$} & $9.054$ & \blue{$83.34\%$} & $8.16$ & \blue{$88.31\%$} & $5.372$ & \blue{$88.16\%$} \\\hline
40 & $6.038$ & \blue{$91.24\%$} & $9.304$ & \blue{$82.75\%$} & $8.508$ & \blue{$87.92\%$} & $5.608$ & \blue{$87.73\%$} \\\hline
50 & $6.15$ & \blue{$90.80\%$} & $9.538$ & \blue{$83.04\%$} & $8.818$ & \blue{$85.87\%$} & $5.782$ & \blue{$86.09\%$} \\\hline
\end{tabular}
\end{table}

\subsection{Computational Complexity Analysis}
\begin{figure}[!ht]
\centering
\includegraphics[width=0.6\columnwidth,keepaspectratio]{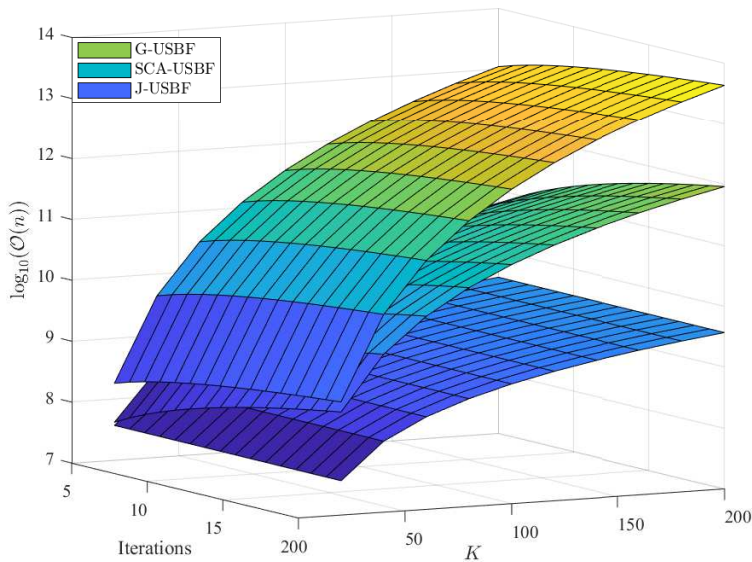}
\captionsetup{labelfont={footnotesize,color=red},font={footnotesize,color=red}}
\color{red}\caption{The floating-point computational magnitude of the algorithms.}
\label{Fig_Complexity}
\end{figure}

\textred{In this subsection, the computational complexity of G-USBF, SCA-USBF and J-USBF is analyzed and compared. Considering the differences in implementation platforms and algorithm design languages, we count the floating-point computation of the proposed algorithms. Firstly, G-USBF includes the US optimization and BF design, whose floating-point computation is about $\sum\limits_{\hat{k}=2}^{K}4(K-\hat{k}+1)(I_{1}(\hat{k}^{3}N+5\hat{k}^{2}N)+\hat{k}^{2})$, where $\hat{k}$ and $I_{1}$ represent the number of scheduled users and iterations, respectively. Secondly, SCA-USBF includes the inner and outer optimizations, whose floating-point computation is about $4I_{3}(I_{2}(7K^{2}N+4KN+14K^{2})+K(N^{3}+2N^{2}+2N))$, where $I_{2}$ and $I_{3}$ represent the number of iterations for both parts. For J-USBF, since the JEEPON model is trained offline, we mainly consider the computation of the testing stage, including the graph representation module, the GCN module and the SINR module. For simplicity, we assume that the GCN module is composed by MLPs with the dimensions $\mathcal{H}\triangleq\{h_{i}\}$. Therefore, the floating-point computation of J-USBF is about $2(2K^{2}N+2KN^{2}+K\sum\limits_{\ell=1}^{L}\sum\limits_{i=1}^{|\mathcal{H}|}(2+h_{\ell,i-1})h_{\ell,i})$. For intuitive comparison, Fig.~\ref{Fig_Complexity} illustrates the comparison of the floating-point computational magnitude of each algorithm for different number of users and iterations. The computational magnitude of J-USBF is lower than that of G-USBF and SCA-USBF, which indicates its computational efficiency advantage.}

\section{Conclusions}
In this paper, the joint US-BF optimization problem is studied for the multiuser MISO downlink system. Specifically, with the help of uplink-downlink duality theory and mathematical transformations, we formulate the original problem into a convex optimization problem, and propose the G-USBF, SCA-USBF and the J-USBF. Numerical results show that J-USBF achieves close performance and higher computational efficiency. Additionally, the proposed J-USBF also enjoys the generalizability in dynamic WCN scenarios. \textred{For future directions, solving the problem of unbearable CSI acquisition burden and signaling overhead caused by the instantaneous perfect CSI applied in this work is interesting and meaningful. Deep learning based resource allocation algorithm needs to be redesigned, and statistical CSI may be helpful to achieve the goal.}

\begin{appendices}

\section{Design of The G-USBF Algorithm}
\textred{In this appendix, the G-USBF algorithm is proposed to slove problem~\eqref{Eq.(05)}, which is inspired by the work in~\cite{zhang2011adaptive} and the near-far effect of WCNs. The feasibility problem in reference~\cite[problem (35)]{he2020beamforming} forms the basis of G-USBF, which is formulated as follows
\begin{subequations}\label{Eq.(A01)}
\begin{align}
\min\limits_{\{\mathbf{w}_{k}\}}\sum\limits_{k\in\mathcal{S}}||\mathbf{w}_{k}||^{2},\\
\mathrm{s.t.}~{r_k}\leq{R}(\overrightarrow{\gamma}_{k}),
\end{align}
\end{subequations}
where $\mathbf{w}_{k}\in\mathbb{C}^{N\times{1}}$ is the BF vector of user $k$, and its downlink power is denoted as $p_{k}=\|\mathbf{w}_{k}\|^{2}$. Solving problem~\eqref{Eq.(A01)} can be used to determine whether the scheduled user set $\mathcal{S}$ is feasible, i.e., the user rate constraint and the BS power budget need to be satisfied. G-USBF is designed with two stages, namely, the conventional greedy search stage and the user set optimization stage, as summarized in \textbf{Algorithm}~\ref{Alg.(A01)}. Here, G-USBF expands the scheduled user set $\mathcal{S}$ from the candidate user set $\mathcal{K}$ in the first stage, and then optimizes $\mathcal{S}$ in the second stage to achieve the goal of scheduling more users. Since the G-USBF algorithm has close performance and lower computational complexity compared with the exhaustive search algorithm, therefore, it is used as the baseline.}


\begin{algorithm}[!ht]
{\color{red}
\caption{The G-USBF Algorithm for Problem~\eqref{Eq.(05)}}\label{Alg.(A01)}
\begin{algorithmic}[1]
\STATE Input candidate user set $\mathcal{K}$ and user CSI $\{\mathbf{h}_{k}\}$, and initialize scheduled user set $\mathcal{S}=\varnothing$.
\STATE Sort the user channels of $\mathcal{K}$ from good to bad via the MRT method, and add the top-ranked user to $\mathcal{S}$.
\STATE In the greedy search stage, move one user from $\mathcal{K}$ to $\mathcal{S}$ in sequence without repetition, and obtain temporary user sets with $|\mathcal{K}\backslash\mathcal{S}|$ groups.
\STATE For each temporary user set, solve problem~\eqref{Eq.(A01)} to obtain $\{p_{k},\mathbf{w}_{k}\}$, and preserve the user set $\mathcal{S}_{1}^{(\ast)}$ with the smallest required power.
\STATE Let $\mathcal{K}\leftarrow\mathcal{K}\backslash\mathcal{S}_{1}^{(\ast)},\mathcal{S}\leftarrow\mathcal{S}_{1}^{(\ast)}$ if $\mathcal{K}\neq\varnothing$ and $\sum\limits_{k\in\mathcal{S}}p_{k}\leq{P}$ is obtained, then go to step 3. Otherwise, go to step 6.
\STATE In the user set optimization stage, move one user with the largest power consumption from $\mathcal{S}$ to $\mathcal{K}$, and obtain the user set $\mathcal{S}_{2}$.
\STATE Let $\mathcal{S}\leftarrow\mathcal{S}_{2}$ and run the greedy search again to obtain a new user set $\mathcal{S}_{2}^{(\ast)}$. If $|\mathcal{S}_{1}^{(\ast)}|=|\mathcal{S}_{2}^{(\ast)}|$, stop iteration then output $\mathcal{S}_{2}^{(\ast)}$ and $\{p_{k},\mathbf{w}_{k}\}$. Otherwise, let $\mathcal{S}\leftarrow\mathcal{S}_{2}^{(\ast)}$ and go to step 6.
\end{algorithmic}}
\end{algorithm}

\section{Design of The CNN-USBF Algorithm}
\textred{In this appendix, the CNN-USBF algorithm is proposed to slove problem~\eqref{Eq.(23)}, which is inspired by the work in~\cite{li2021survey}. In particular, CNN-USBF takes the WCN graph representation as input and outputs the US-PA optimization strategy and BF vectors. To be specific, the update rule of CNN-USBF for node $v$ in graph $\mathcal{G}(\mathcal{V},\mathcal{E})$ is formulated as}
{\color{red}\begin{equation}\label{Eq.(32)}
\begin{aligned}
\mathrm{Input:}&~\mathbf{D}_{v}^{(0)}=[\mathbf{x}_{v},F_{\mathrm{max}}(\{\mathbf{e}_{u,v}\}),F_{\mathrm{mean}}(\{\mathbf{e}_{u,v}\})],u\in\mathcal{N}_{v},\\
\mathrm{CNN\raisebox{0mm}{-}stage:}&~\mathbf{D}_{v}^{(i)}=F_{\mathrm{std}}(\mathrm{Cov1d}(\mathbf{D}_{v}^{(i-1)})),i=1,2,\cdots,N_{\mathrm{1}},\\
\mathrm{DNN\raisebox{0mm}{-}stage:}&~\mathbf{D}_{v}^{(i)}=F_{\mathrm{std}}(\mathrm{LNN}(\mathbf{D}_{v}^{(i-1)})),i=N_{\mathrm{1}}+1,\cdots,N_{\mathrm{1}}+N_{\mathrm{2}},\\
\mathrm{Output:}&~\mathbf{D}_{v}^{(N_{\mathrm{2}})}=[\kappa_{v}^{(\ast)},q_{v}^{(\ast)}], \mathrm{and~BF~vector}~\mathbf{w}_{v}^{(\ast)},v\in\mathcal{V},
\end{aligned}
\end{equation}}
\textred{where $N_{\mathrm{1}}$ and $N_{\mathrm{2}}$ denote the layers of CNN and DNN, respectively. $\mathbf{D}_{v}^{(0)}$ is the features of node $v$ and its neighborhood edges, $\mathbf{D}_{v}^{(N_{\mathrm{2}})}$ is the US-PA strategy of node $v$, and $F_{\mathrm{std}}(\mathbf{z})=F_{\mathrm{AC}}(F_{\mathrm{BN}}(\mathbf{z}))$ is the standardization function used to standardize the network input to accelerate training process and reduce generalization error, which is implemented by BN and AC layers. The neural network module of CNN-USBF is constructed through CNN and DNN, which are implemented and trained by \emph{Pytorch} and PDLF, respectively. The algorithm steps of CNN-USBF refer to J-USBF. Note that unless mentioned otherwise, the neural network structure of CNN-USBF refer to Table~\ref{Tab-08} and is trained separately for different WCN scenarios.}

\begin{table}[!ht]
\centering
\renewcommand{\arraystretch}{1.1}
\captionsetup{labelfont={color=red},font={color=red}}
\caption{The neural network structure of CNN-USBF.}\label{Tab-08}
{\color{red}\begin{tabular}{|l|l|}
\hline
Layer & Parameters \\\hline
Layer 1 (Input) & Input of size $3K$, batch of size $K$, $N_{\mathrm{e}}$ epochs \\\hline
Layer 2 (Cov1d, BN and AC) & Input=$3$, output=$256$; Input=$256$; LReLU \\\hline
Layer 3 (Cov1d, BN and AC) & Input=$256$, output=$128$; Input=$128$; LReLU \\\hline
Layer 4 (Cov1d, BN and AC) & Input=$128$, output=$64$; Input=$64$; LReLU \\\hline
Layer 5 (LNN, BN and AC) & Input=$64$, output=$32$; Input=$32$; LReLU \\\hline
Layer 6 (LNN, BN and AC) & Input=$32$, output=$16$; Input=$16$; LReLU \\\hline
Layer 7 (LNN, BN and AC) & Input=$16$, output=$2$; Input=$2$; LReLU \\\hline
Layer 8 (Output and PAC) & Output of size $2K+KN$, Adam optimizer \\\hline
\end{tabular}}
\end{table}

\end{appendices}

\begin{small}

\end{small}
\end{document}